\documentclass{article}

\usepackage{arxiv}

\usepackage[utf8]{inputenc} % allow utf-8 input
\usepackage[T1]{fontenc}    % use 8-bit T1 fonts
\usepackage{hyperref}       % hyperlinks
\usepackage{url}            % simple URL typesetting
\usepackage{booktabs}       % professional-quality tables
\usepackage{amsfonts}       % blackboard math symbols
\usepackage{nicefrac}       % compact symbols for 1/2, etc.
\usepackage{microtype}      % microtypography
\usepackage{lipsum}

\usepackage{graphicx}
\usepackage{tabularx}
\usepackage{multirow}
\usepackage{color,array,dcolumn}
\usepackage{amsmath}
\usepackage{bbold}
\usepackage{multirow}
\usepackage{subcaption}

\newcolumntype{L}[1]{>{\raggedright\let\newline\\\arraybackslash\hspace{0pt}}m{#1}}
\newcolumntype{C}[1]{>{\centering\let\newline\\\arraybackslash\hspace{0pt}}m{#1}}
\newcolumntype{R}[1]{>{\raggedleft\let\newline\\\arraybackslash\hspace{0pt}}m{#1}}

\usepackage[inline]{trackchanges} %for better track changes. finalnew option will compile document with changes incorporated.
\addeditor{TODO}
\addeditor{OP}
\addeditor{MS}
\addeditor{VM}

\newcommand{\x}{\mathbf{x}}
\newcommand{\y}{\mathbf{y}}
\newcommand{\E}[1]{\mathbb{E}\left[#1\right]}
\newcommand{\mP}{\mathbf{P}}
\newcommand{\mM}{\mathbf{M}}
\newcommand{\mK}{\mathbf{K}}
\newcommand{\mR}{\mathbf{R}}
\newcommand{\mH}{\mathbf{H}}
\newcommand{\mI}{\mathbf{I}}
\newcommand{\mF}{\mathbf{F}}
\newcommand{\mQ}{\mathbf{Q}}

\newcommand{\Eq}[1]{Eq.~\eqref{#1}}
\newcommand{\Fig}[1]{Fig.~\ref{#1}}

\newcommand{\ie}{\textit{i.e.} }
\newcommand{\eg}{\textit{e.g.} }

\usepackage{time}

\title{Boundary Conditions for the Parametric Kalman Filter forecast\\\small{version of: \today, \now}}

 \author{
   M. Sabathier\thanks{Corresponding author} \\
   ONERA, Toulouse, France\\
   \texttt{martin.sabathier@onera.fr} \\
    \And
   O. Pannekoucke \\
   INPT-ENM, Toulouse, France\\
CNRM, Université de Toulouse, Météo-France, CNRS, Toulouse, France\\
CERFACS, Toulouse, France\\
   \texttt{olivier.pannekoucke@meteo.fr} \\
    \And
   V. Maget\\
   ONERA, Toulouse, France\\
   \texttt{vincent.maget@onera.fr} \\
   \And
   N. Dahmen\\
   ONERA, Toulouse, France\\
   \texttt{nourallah.dahmen@onera.fr} \\ 
}

\begin{document}

\maketitle

% \authors{M.~Sabathier\affil{1}, O.~Pannekoucke\affil{2,3,4}, V.~Maget\affil{1}, N.~Dahmen\affil{1}}

% \affiliation{1}{ONERA, Toulouse, France}
% \affiliation{2}{CNRM, Université de Toulouse, Météo-France, CNRS, Toulouse, France}
% \affiliation{3}{CERFACS, Toulouse, France}
% \affiliation{4}{INPT-ENM, Toulouse, France}

%\correspondingauthor{Martin Sabathier}{martin.sabathier@onera.fr}

\begin{abstract}
This paper is a contribution to the exploration of the parametric Kalman filter (PKF), which is an approximation of the Kalman filter, where the error covariances are approximated by a covariance model. Here we focus on the covariance model parameterized from the variance and the anisotropy of the local correlations, and whose parameters dynamics provides a proxy for the full error-covariance dynamics. For this covariance model, we aim to provide the boundary condition to specify in the prediction of PKF for bounded domains, focusing on Dirichlet and Neumann conditions when they are prescribed for the physical dynamics. An ensemble validation is proposed for the transport equation and for the heterogeneous diffusion equation over a bounded 1D domain. This ensemble validation requires to specify the auto-correlation time-scale needed to populate boundary perturbation that leads to prescribed uncertainty characteristics. The numerical simulations show that the PKF is able to reproduce the uncertainty diagnosed from the ensemble of forecast appropriately perturbed on the boundaries, which show the ability of the PKF to handle boundaries in the prediction of the uncertainties. It results that Dirichlet condition on the physical dynamics implies Dirichlet condition on the variance and on the anisotropy.
\end{abstract}


\section*{Plain Language Summary}
This work addresses the question of the uncertainty prediction in bounded domains. It contributes to explore a theoretical formulation of the uncertainty prediction, the parametric Kalman filter, that opens the way to data assimilation in real applications where the boundaries are important, such as in radiation belts predictions, air quality, or wild-land fire ; or the exploration of the coupling of the uncertainty in the atmosphere-ocean coupled system.

\section{Introduction}

Uncertainty prediction is a challenging topic, important in probabilistic forecasting as well as in data assimilation (DA) \eg in variational data assimilation to improve the specification of the background term \cite{Lorenc2003QJRMS,Buehner2005QJRMS}. One of the main theoretical backbone in DA is given by the Kalman filter equations that applies for linear dynamics, but that cannot be used for large systems because the numerical cost to predict the error covariance becomes prohibitive. Hence, approximations of the KF have been proposed, as the ensemble Kalman filter (EnKF) where the error covariance matrices are approximated from ensemble estimation \cite{Evensen2009book}.

The EnKF is widely used and has shown to perform well for many applications in geosciences \eg for the weather prediction \cite{Houtekamer2001MWR}, or for the radiation belts prediction \cite{Bourdarie2012AG}. Radiation belts dynamics modeling consist in estimating quantitatively the fluxes of high energetic electrons and protons trapped in the Earth magnetic field using a typical advection-diffusion equation. This region spans from 1 Earth Radius up to 8, thus encompassing all typical satellites orbits, with which such particles can strongly interact and induce minor to critical onboard anomalies. Compared with global prediction, radiation belts predictions are performed on a limited and non-periodic domain where the boundary imposes conditions to the dynamics of the electrons and protons. Indeed, on one side, the outer boundary condition is considered as the prime access for fresh materials, coming from the so-called magneto-tail, and is typically modeled as an imposed Dirichlet condition at this altitude (8 Earth radii) that can evolve as a function of solar activity (e.g. energy spectrum reshaping from time to time) \cite{Maget2015JGR}. On the other side, close to the Earth, the atmosphere implies a necessary fixed Dirichlet condition too, as all radiation belts particles coming down there are absorbed (e.g. distribution always equal to 0). Finally, for low energy boundary we expect to rely on a Neumann condition to limit naturally any escape of particles or artificial source. Nonetheless, when performed on limited area models, atmospheric prediction also present such kind of boundaries.

In weather forecasting over local-area domains, the importance of the boundary has been identified from sensitivity studies showing that errors may be traced back to the boundaries \cite{Errico1993TAa}. Hence, inaccurate boundary conditions may result in large forecast errors \cite{Vukicevic1990MWR,Vukicevic1989MWR,Warner1989MWR}. 
However, in preliminary studies on the potential of EnKF applied to local-area models (LAM), no perturbations of the boundary conditions were considered \cite{Snyder2003MWRa,Dowell2004MWR}, leading to a loss of variance as lead time increases \cite{Nutter2004MWR,Nutter2004MWRa}.
Note that, this loss of variance also occurs for the Kalman filter prediction step \textit{e.g.} in a transport equation over 1D limited-area domain, when the boundary condition is not well formalized in the dynamics, the inflow of variance is null at the inlet of the domain (see \ref{AppKFBC}). This issue can be solved from the formalism of the Kalman filter as used in optimal control, by considering that the boundary is an uncertain command as done in \eg heat conduction estimation \cite{Scarpa1995NHT}.
In the community of the EnKF for LAMs, the loss of variance is avoided by perturbing the boundary conditions \cite{Nutter2004MWRa}, \textit{e.g.} 
by modeling the spatial and temporal covariance relationship of the boundary condition based on the use of multivariate covariance model 
or by using boundary conditions which derives from an ensemble on a larger domain \cite{Torn2006MWR}, even if for the latter approach, the consistency across multiple domains is difficult to handle \cite[sec. 6.a]{Houtekamer2016MWR}.
Note that in variational data assimilation for LAMs, it is possible to include the uncertainty on the boundary 4DVar through the adjunction of a cost function term that constraints the increment or its tendency at the boundary \cite{Gustafsson2012TA}, which is a way to avoid the loss of variance above mentioned for the EnKF and for the KF. 

In radiation belts where there is no larger model and thus no ensemble  from which perturbation on the boundary can be introduced, we are interested by sampling boundary perturbations from a spatio-temporal covariance model. However, because the solar activity is non-stationary, we need to develop perturbations of the boundary conditions that are  heterogeneous and non-stationary \ie with varying time auto-correlation, in place to the auto-regressive model often used but for which the time auto-correlation is constant \cite{Torn2006MWR}.
Moreover, we want to reduce the numerical cost of using an EnKF while keeping the uncertainty dynamics it provides.
This motivation can be also be encountered in the development of the assimilation scheme for operational model in air quality, that deals with hundreds of chemical species, and for which the numerical cost of an integration is large. For instance, in the operational air quality model MOCAGE developed at M\'et\'eo-France, the assimilation relies on a 3D-FGAT approach \cite{Amraoui2020AMT,Massart2010MWR}, without ensemble, and it would be interesting to introduce flow-dependency of the background term while accounting for the transport of the boundary uncertainty during the assimilation at regional scale.

Recently another approximation for the KF, different from the EnKF, has been introduced, the parametric Kalman filter (PKF), where the error covariance matrices are approximated from a covariance model \cite{Pannekoucke2016T}. In the PKF, the dynamics of the parameters provides a proxy for the dynamics of the full covariance matrix. For instance, covariance model parameterized from the variance and the anisotropy of the local correlation functions are able to predict the dynamics of the covariance matrix for transport equations \cite{Cohn1993MWR}, but at a numerical cost equivalent to three time the integration of the transport \cite{Pannekoucke2018NPG,Pannekoucke2021T}. 
In addition, the PKF provides an understanding of the uncertainties dynamics, which is less obvious to determine from an ensemble estimation alone. For instance, when considering a transport, the PKF equations for the error statistics consist in:  the advection of the error variance without source term, meaning that the magnitude of the error variance is conserved ;  and  the advection of the anisotropy plus a source term that corresponds to the deformation of the anisotropy due to the local shear \cite{Pannekoucke2021T}. As another example, when considering now a  conservative dynamics or a non-linear Burgers dynamics (in the tangent-linear approximation for the error statistics), a source term appears in the PKF equation for the error variance which is related to high gradient area that can be interpreted as a production term similar to the one encountered in turbulence \cite{Pannekoucke2018NPG}. These examples of physical interpretations of the dynamics of the error statistics, as it is provided by the PKF equations, are less easy to deduce from the ensemble estimation while the effects of these processes can be observed. Note that data-driven modeling leveraged on deep-learning, especially physics-informed neural network \cite{Raissi2019JCP}, can be considered to determine the dynamics of the error statistics \cite{Pannekoucke2020GMD}. On the other hand, it is not necessarily easy to interpret all terms in the PKF equations \eg when considering a diffusion equation, for which the PKF shows a coupling between the error variance and the anisotropy, it is hard to physically explain terms in square of the gradient of the anisotropy.
In this way, the PKF can contribute to improve the understanding of other topics of interest in data assimilation \eg in the characterization of the dynamics of the model-error covariance due to the spatial and temporal discretization of a transport equation \cite{Pannekoucke2021NPG,Menard2021QJRMS}. The PKF is interesting to provide a flow dependency of the background error covariance matrix in variational data assimilation \eg by providing a proxy for the diffusion tensor of the day in the covariance model based on the diffusion equation \cite{Pannekoucke2016T} ;
or by providing the variance of the day in presence of model error thanks to the PKF dynamics of the variance, as in the assimilation of GOSAT methane observations using the hemispherical CMAQ \cite{Voshtani2022RS}. While the PKF has been mainly developed in univariate statistics, an exploration has begun to develop a multivariate statistics formulation \cite{Perrot2022}. At a more theoretical level, the parametric formulation has been considered to describe the covariance dynamics of white noise error in a conservative equation \cite{Gilpin2021A}.
Hence, the PKF approach seems promising to tackle the issues encountered in the development of the data assimilation for radiation belts and air quality.

Until now, the PKF has been explored mainly on periodic 1D or 2D domains, where it has been shown to reproduce interesting features of the uncertainty dynamics in linear problem \eg for the transport \cite{Pannekoucke2021T}, as well as for non-linear dynamics at the second order \eg for the non-linear advection-diffusion equation \cite{Pannekoucke2018NPG}. Note that the non-linear extension of the PKF relies on a tangent-linear evolution of error, with a feedback of the uncertainty on the dynamics of the mean through the error-variance (\textit{i.e.} a fluctuation-mean interaction), leading to a Gaussian second-order filter \cite[Sec. 9.3]{Jazwinski1970book}.
However, to go ahead towards real applications, and especially applications in bounded domains, appropriate specification of boundary conditions of the error statistics is needed for the PKF dynamics. To do so, we propose to explore the specification of the boundary conditions for the PKF when Dirichlet and Neumann conditions are considered in the physical dynamics. This exploration focuses on two dynamics of interest for our applications: the transport equation \eg for air quality or weather prediction ; and the diffusion equation \eg for radiation belts prediction or uncertainty dynamics in boundary layer for air quality. The paper focuses on the forecast step, and the assimilation step is not addressed here.

The paper is organized as follows. First, the background of the parametric Kalman filter (PKF) is reminded in Section~\ref{sec:reminder_KF_PKF}. Then, Section~\ref{sec:pkfbc} details how to specify the PKF conditions at the boundary for the forecast for the Dirichlet and the Neumann conditions. The ensemble validation of the boundary conditions for the PKF needs an ensemble of forecasts. To do so, an intermediate Section~\ref{sec:EnKF:bc} will details how to specify boundary conditions in an EnKF experiment that produces desired error statistics. This is an important contribution of the paper in order to validate the specification of the boundary conditions of the PKF, where the numerical validation is presented in Section~\ref{sec:num:validation}. Conclusions and perspectives are given in the last Section~\ref{sec:conclusion}.

\section{Background on the PKF forecast step} \label{sec:reminder_KF_PKF}

This section gives a self contained introduction to the parametric Kalman filter, applied for a particular covariance model. First, the prediction step of the Kalman filter applied on a linear dynamics is reminded. Then, the formalism of the PKF is presented, followed by the illustration on two dynamics: the transport equation, important in geosciences, and the diffusion equation important in radiation belt dynamics community.

\subsection{Kalman filter forecast step}

Here we consider the prediction of a univariate physical field $\chi(t,\x)$ defined on a domain $\Omega$ of dimension $d$ and coordinate system $\x=(x^i)_{i\in[1,d]}$, whose dynamics is given by 
\begin{equation}\label{eq:dynamics}
    \partial_t \chi = \mathcal{M}(\partial\chi),
\end{equation}
where $\mathcal{M}$ stands for a function of the state $\chi$ and of its spatial derivatives, $\partial\chi$, which is a shorthand for the partial derivative with respect to the spatial coordinates at any arbitrary orders, with the convention that order zero denotes the field $\chi$ itself.
Thereafter, for the sake of simplicity, $\mathcal{M}$ is assumed linear but the formalism extends to the non-linear framework \cite{Pannekoucke2018NPG,Pannekoucke2021GMD}. Note that $\chi$ can be either continuous or discrete (the discretized version of the continuous field): the discrete case leads to matrix algebra relations \eg $\mathcal{M}$ is replaced by its matrix formulation $\mM$. 

Hence, for a time integration over a time window $[0,T]$, the dynamics \Eq{eq:dynamics} is written as 
\begin{equation}\label{eq:KFforecast}
    \chi(T)=\mM_{T\leftarrow 0}\chi(0),
\end{equation} 
where $\mM_{T\leftarrow 0}$ denotes the propagator associated with the time integration of \Eq{eq:dynamics} over  $[0,T]$. 
Compared to the classical formulation of the Kalman filter equations, we do not consider here the model error because it will be made negligible in the numerical experiments performed in this study by considering a sufficiently fine space-time resolution.

In real applications, the spatio-temporal heterogeneity of the observation network, as well as the model error, imply that $\chi$ is not known exactly. The true state of the system is denoted by $\chi^t$. The analysis state, that is the estimation of the true state knowing the observations until a given time, is denoted by $\chi^a$. The deviation of the analysis state from the truth is the analysis error, $e^a=\chi^a-\chi^t$, and is often modeled as a random Gaussian vector of zero mean and covariance matrix $\mP^a=\E{e^a(e^a)^\mathrm{T}}$, where $\E{\cdot}$ stands for the expectation operator and where the upper script $(\cdot)^\mathrm{T}$ stands for the transpose operator (later the adjoin operator for matrices). Hence, the distribution of the initial condition is a Gaussian of mean $\chi^a$ and covariance matrix $\mP^a$, denoted by $\mathcal{N}\left(\chi^a,\mP^a\right)$. The forecast state at a time $T$, $\chi^f(T)=\mM_{T\leftarrow 0}\chi^a$ provides an approximation of the true state at time $T$. For linear dynamics and Gaussian uncertainty, the forecast error $e^f(T)=\chi^f(T)-\chi^t(T)$ is a Gaussian vector of zero mean and covariance matrix $\mP^f(T)=\E{e^f(e^f)^\mathrm{T}}(T)$, whose dynamics is written as 
\begin{equation}\label{eq:dyne}
    \partial_t e^f=\mathcal{M}(\partial e^f).
\end{equation}

The forecast-error covariance matrix is related to the analysis-error covariance matrix by 
\begin{equation}\label{eq:KF}
    \mP^f(T) = \mM_{T\leftarrow 0}\mP^a\left(\mM_{T\leftarrow 0}\right)^\mathrm{T}.
\end{equation}
Equation \eqref{eq:KF} corresponds to the Kalman filter propagator of the error covariance matrix, whose the particular dynamics is given by 
\begin{equation}\label{eq:dynKF}
    \frac{d\mP^f}{dt} = \mM\mP^f + \mP^f\mM^\mathrm{T},
\end{equation}
integrated over the period $[0,T]$, starting from the initial condition $\mP^f(0)=\mP^a$. Hence, the distribution of the forecast at $T$ is a Gaussian of mean $\chi^f(T)$ and covariance matrix $\mP^f(T)$, that is $\mathcal{N}\left(\chi^f,\mP^f(T)\right)$.

When observations, $y^o$, are available at $T$, so that $y^o = \mH \chi^t(T)+e^o$ where $\mH$ denotes the linear observation operator that maps a state (here the true state $\chi^t(T)$) into the observation space, and $e^o$ denotes the observational error, modeled as a Gaussian random vector of zero mean and covariance matrix $\mR$ and assumed uncorrelated from the forecast error $e^f(T)$ ; then the Kalman analysis equations at time $T$ are written as 
\begin{subequations}
    \label{eq:kf:analysis}
    \begin{align}
        \chi^a(T) &= \chi^f(T) + \mK(y^o-\mH \chi^f(T)),\label{eq:kf:analysis:a}\\
        \mP^a(T) &= (\mI-\mK\mH)\mP^f(T),\label{eq:kf:analysis:b}
    \end{align}       
\end{subequations}
where $\mK=\mP^f(T)\mH^T(\mH\mP^f(T)\mH^T+\mR)^{-1}$ is the Kalman gain matrix. \Eq{eq:kf:analysis} characterizes the Gaussian distribution of the analysis uncertainty, where \Eq{eq:kf:analysis:a} is the equation of the mean, and \Eq{eq:kf:analysis:b} is the equation of the analysis-error covariance matrix at time $T$. \Eq{eq:KF} together with \Eq{eq:kf:analysis} are the Kalman filter equations, that can be cycled providing the analysis and forecast assimilation cycles, where \Eq{eq:KF} (\Eq{eq:kf:analysis}) is the KF forecast (analysis) step.

Note that this contribution being focused on the uncertainties dynamics, with the corresponding Kalman filter forecast of the covariance matrix \Eq{eq:KF}, the uncertainty related to the observations (present when considering cycled assimilations) appears only through the statistics of the analysis error, and more precisely through the covariance matrix of the analysis error $\mP^a=\mP^f(0)$. 
Moreover, since we consider the model as perfect, the uncertainty considered here only comes from the initial condition and the boundary conditions.

Note also that the derivation of the Kalman filter forecast step \Eq{eq:KF} follows the usual one encountered in the weather community, and can leads to drawback when applied to LAMs without care \textit{e.g.} a loss of variance at the inlet of the domain. We detail this issue considering a transport in a 1D limited-area domain in \ref{AppKFBC}. However, it results that when boundary conditions are uncertain, it is necessary to reformulate the problem. That can be done in the KF framework, by considering that the BCs are an uncertain command following the terminology used in optimal control (see \eg \cite{Scarpa1995NHT}).

\subsection{Ensemble approximation of the Kalman filter forecast step}

While the KF forecast step \Eq{eq:KF} is a simple algebraic formula, it fails to apply in large systems because of its numerical cost: if $n$ denotes the dimension of the vector representation of $\chi$, then the computational complexity of \Eq{eq:KF} scales between $n^2$ and $n^3$ \cite{Strassen1969NM}. In terms of integration cost, the KF requires $2n$ integrations of the model \Eq{eq:dynamics}.

Hence, approximations for the KF are needed. For instance, in the Ensemble Kalman filter (EnKF), the forecast error-covariance matrix is approximated by its ensemble estimation.
\begin{equation}\label{eq:Pfe}
    \widehat{\mP^f}(t)=\frac{1}{N}\sum_k e_k e_k^\mathrm{T},
\end{equation}
with $e_k=\chi_k(t)-\widehat{\chi}(t)$ where
$\widehat{\chi}(t)=\frac{1}{N}\sum_k \chi_k(t)$ denotes the empirical mean and $\left(\chi_k(t)\right)_{k\in[1,N]}$ is an ensemble of $N$ forecasts \cite{Evensen2009book}. This time, the numerical complexity scales with the number of ensemble members $N$ and the size of the problem $n$: the numerical cost of an ensemble of forecast is the cost of $N$ integrations of the model \Eq{eq:dynamics}.

Note that the normalization by $N$ in \Eq{eq:Pfe} leads to a bias that decreases as $1/N$. In EnKF framework, the normalization by $N-1$ is preferred, however since we latter consider estimation from very large ensemble size, the corrections of the estimators are not considered here, and we only consider empirical mean estimations $\frac{1}{N}\sum_k(\cdots)$ as in \Eq{eq:Pfe}.

In the ensemble Kalman filter forecast step applied for LAMs, the uncertainty on the boundary condition can be introduced by adding perturbations at the boundary for each member of the ensemble. This can be done from considering a modeling of the multivariate covariance at the boundary or by considering boundary condition derived from an ensemble on a larger domain \cite{Torn2006MWR}.

While this contribution focuses on the forecast step, we can mention that if ensemble computation of the KF analysis step \Eq{eq:kf:analysis} can be implemented following different algorithm \eg the EnKF version based on perturbation of observation introduced by \cite{Burgers1998MWR}, the square root filter of \cite{Tippett2003MWR} or the ETKF of \cite{Bishop2001MWR} and its variational implementation \cite{Harlim2007TA,Bocquet2011NPG} ; the KF forecast step is mainly the same whatever the variant of the EnKF considered. Hence, for this study dedicated to the exploration of the specification of the boundary conditions for the PKF forecast step, the particular implementation the EnKF has not impact and no assimilation experiment is conducted here.

The next section presents another approximation for the error-covariance matrices.

\subsection{Parametric formulation for the Kalman filter forecast step based on VLATcov models}

In the parametric approach, a covariance model is introduced, $\mP(\mathcal{P})$ where $\mathcal{P}$ denotes the set of parameters of the covariance model, so to approximate the error covariance matrices. For instance, the forecast-error covariance matrix $\mP^f$, is approximated as $\mP(\mathcal{P}^f)\approx\mP^f$, where $\mathcal{P}^f$ is a particular set of values for the parameters.
The parametric Kalman filter (PKF) dynamics aims to mimic the dynamics of \Eq{eq:dynKF} relying on the dynamics of the parameters $\mathcal{P}^f$,
\begin{equation}\label{eq:dynPKF}
    \frac{d\mathcal{P}^f}{dt} = \mathcal{G}(\mathcal{P}^f),
\end{equation}
where $\mathcal{G}$ has to be determined from the particular dynamics of \Eq{eq:dynamics}, so that at any time $t$, $\mP(\mathcal{P}^f(t))$ approximates $\mP^f(t)$ \ie $\mP(\mathcal{P}^f(t))\approx\mP^f(t)$. As for the EnKF, the numerical complexity of the PKF prediction \Eq{eq:dynPKF} scales as number of parameters and the dimension of the problem $n$: the numerical cost of the PKF represent the cost of few numerical integrations of the dynamics \Eq{eq:dynamics}, depending on the number of parameters needed for the covariance approximation.

Thereafter, since we deal with the forecast step of the PKF, the upper-script $^f$ is dropped in the notation that concerns the forecast-error statistics.

This contribution will focus on the particular class of covariance model, so-called VLATcov models, parameterized from two fields, defined below: the variance field, $V$, and the local anisotropy tensor of the correlation functions, $\mathbf{g}$ or $\mathbf{s}$. Hence, the set of parameters is given by the couple $\mathcal{P}=(V,\mathbf{g})$ or $\mathcal{P}=(V,\mathbf{s})$, so that a VLATcov model is written as $\mP(V,\mathbf{g})$ or $\mP(V,\mathbf{s})$.
For an error field $e$, the variance field is defined as 
\begin{equation}\label{eq:V}
    V=\E{e^2},
\end{equation}
% and is used to introduce the normalized error $\varepsilon = \frac{e}{\sqrt{V}}$.
When the error field is a differential random field, that is assumed from now, the correlation function $\rho(\x,\y)=\E{\varepsilon(\x)\varepsilon(\y)}$ is flat for $\y=\x$. Then, the local anisotropy at $\x$ is defined as the local metric tensor $\mathbf{g}(\x)$ (also denoted by $\mathbf{g}_\x$) which appears in the second-order Taylor's expansion
\begin{equation}
    \rho(\x,\x+\delta\x) \approx 1 - \frac{1}{2}||\delta\x||_{\mathbf{g}_\x}^2,
\end{equation}
where 
$||\delta\x||_{\mathbf{g}_\x}^2=\delta \x^\mathrm{T} \mathbf{g}_\x \delta \x $ denotes the norm associated with the metric tensor $\mathbf{g}_\x$ that is the symmetric definite positive matrix $[\mathbf{g}_\x]_{ij}=-\partial_{x^i x^j}^2\rho_\x$
where  $\rho_\x(\delta\x)$ stands for the local correlation function. In a 1D domain of coordinate $x$, the metric tensor field is the scalar field $\mathbf{g}=[g_{xx}]$.

In practice, the geometry of the local metric tensor is
contravariant: the direction of largest correlation anisotropy corresponds to the principal axes of smallest eigenvalue for the metric tensor. Thus, it is useful to introduce
the local aspect tensor \cite{Purser2003} 
\begin{equation}
\label{eq:s}
\mathbf{s}(\x) = \left( \mathbf{g} (\x) \right)^{-1},
\end{equation}
where the superscript $(\cdot)^{-1}$ denotes the matrix inverse, and whose the geometry goes as the correlation.

What makes the local metric tensor attractive is that this tensor is related to the normalized error by (see \eg \cite{Pannekoucke2021T})
\begin{equation}\label{eq:g}
 [\mathbf{g}_\x]_{ij}=\E{\partial_{x^i}\varepsilon\partial_{x^j}\varepsilon}.
\end{equation}
Hence, the variance \Eq{eq:V} and the anisotropy \Eq{eq:g} can be computed from an ensemble estimation: the variance field is estimated by
\begin{equation}
    \label{eq:Ve}
    \widehat{V} = \frac{1}{N}\sum_k \left(e_k(t)\right)^2,
\end{equation}
with $e_k(t)=\chi_k(t)-\widehat{\chi}(t)$, 
from which derivatives of the normalized error $\varepsilon_k=\frac{1}{\sqrt{V}}\left(\chi_k(t)-\widehat{\chi}(t)\right)$ 
leads to the estimation of the upper triangular components of the metric
\begin{equation}
    \label{eq:ge}
    \widehat{g}_{ij} = \frac{1}{N}\sum_k \partial_{x^i}\varepsilon_k \partial_{x^j}\varepsilon_k,
\end{equation}
for $i\leq j$ (since $g_{ji}=g_{ij}$).
While the PKF approach does not relies on any ensembles, the ensemble estimations 
\Eq{eq:Ve} and \Eq{eq:ge} can be used to set the initial conditions for the parameters to ignite the assimilation cycles, or to validate the PKF from the diagnosis of an EnKF. 

An example VLATcov model is given by the heterogeneous Gaussian-like covariance model
\cite{Paciorek2006E}
    \begin{equation}\label{eq:hegauss}
        \mP(V,\mathbf{s})(\mathbf{x},\mathbf{y}) =
    \sqrt{ V_{\mathbf{x}} V_{\mathbf{y}}}
    \frac{|\mathbf{s_x}|^{1/4} |\mathbf{s_y}|^{1/4} }
    {|\frac{1}{2} (\mathbf{s_x} + \mathbf{s_y})|^{1/2}} 
    \exp \left( 
    -\frac{1}{2}||\mathbf{x}-\mathbf{y}||^2_{[\frac{1}{2}(\mathbf{s_x}+\mathbf{s_y})]^{-1}}
    \right)
    \end{equation}    
where $|\cdot|$ denotes the matrix determinant.

When VLATcov models are used for the parametric approach, the dynamics of the parameters \Eq{eq:dynPKF} is deduced from the time derivative of \Eq{eq:V} and \Eq{eq:g}, and the dynamics of the error \Eq{eq:dyne}. 
Hence, in univariate statistics, the PKF equations for VLATcov models consists in three equations. The first equation corresponds to the prediction of the mean, that is \Eq{eq:dynamics} in the linear case.
% Dynamic of the variance
The second equation is the dynamics of the variance, $V$, which is deduced from 
\begin{subequations}\label{eq:dynV}
\begin{equation}\label{eq:dynVa}
    \partial_t V = 2\E{e\partial_t e},
\end{equation}
where replacing the trend of the error \Eq{eq:dyne}, will leads to the dynamics of $V$
\begin{equation}\label{eq:dynVb}
    \partial_t V = 2\E{e\mathcal{M}(\partial e)}.
\end{equation}
\end{subequations}
The third equation is the dynamics of the anisotropy, $g$, 
whose components evolve as 
\begin{equation}\label{eq:dyng}
    \partial_t g_{ij} = \E{\partial_t\left(\partial_{x^i}\varepsilon\partial_{x^j} \varepsilon\right)},
\end{equation}
and from which we can deduce the dynamics of the aspect tensor $\mathbf{s}=\mathbf{g}^{-1}$ with
$\partial_t \mathbf{s} = -\mathbf{s}(\partial_t \mathbf{g})\mathbf{s}$.
The dynamics of the variance \Eq{eq:dynVb} and of the anisotropy \Eq{eq:dyng} can be simplified \eg by considering the commutation between the expectation and partial derivatives \cite{Pannekoucke2021GMD}.

Note that for non-linear dynamics, the PKF forecast step for VLATcov models stands as (see \cite{Pannekoucke2021GMD} for details)
\begin{subequations}
\label{eq:pkf:fcst}
\begin{align}
\partial_t \E{\chi} &= \mathcal{M}(\partial \E{\chi}) + \mathcal{M}''(\partial \E{\chi})\left(\E{\partial e \otimes \partial e}\right),\label{eq:pkf:fcst:a}\\
\partial_t V &= 2\E{e\partial_t e},\label{eq:pkf:fcst:b}\\
\partial_t g_{ij} &= \E{\partial_t\left(\partial_{x^i}\varepsilon\partial_{x^j} \varepsilon\right)},\label{eq:pkf:fcst:c}
\end{align}
\end{subequations}
where 
$\mathcal{M}'$ and $ \mathcal{M}''$ with two linear operators : the former (the
latter) refers to the tangent-linear model (the Hessian), and
both are computed with respect to the mean state $\E{\chi}$ ; 
$\partial e \otimes \partial e$ denotes the tensor product of the partial derivatives with respect to the spatial coordinates, i.e., terms such
as $\partial^k e \partial^m e$ for any positive integers $(k, m)$ ; and where the trends than appears in the right-hand side of \Eq{eq:pkf:fcst} should be formulated from the dynamics \eg using $\partial_t e = \mathcal{M}'(\partial \E{\chi})(\partial e)$ for the trend of $e$.

The computation of dynamical equations \Eq{eq:pkf:fcst} for the mean, the variance $V$ and  the anisotropy $\mathbf{g}$ (or $\mathbf{s}$) can be performed using a computed algebra system. To do so, the open source Python toolbox SymPKF has been introduced  \cite{git-symPKF,Pannekoucke2021GMD}, which computes the dynamics of the parameters and renders a numerical code to facilitate the numerical exploration of the PKF approach. Another way to simplify the computation of the parameters dynamics is to identify the contribution of each physical process present in \Eq{eq:dynamics} following a splitting strategy \cite{Pannekoucke2018NPG,Pannekoucke2021GMD}. Thereafter, the dynamics of the VLATcov parameters is computed by using SymPKF and the interested reader is referred to the Jupyter notebooks that are provided as a supplementary material to this contribution \cite{git-pkf-boundary}.

Compared to the ensemble implementation of the KF (e.g. the EnKF or the ETKF), the contribution of the PKF  is that it replaces the computational cost for the ensemble prediction step by the numerical cost of the time integration of the system \Eq{eq:pkf:fcst}. More precisely, in univariate statistics, the PKF based on the VLATcov model scales as the number of independent components in $\mathbf{g}$ (the number of coefficients in the upper triangle) plus one for the variance field: in a univariate over a 1D (3D) domain, this represents 2 (7) times the cost of one model forecast (which scales itself with the dimension $n$). This numerical cost of the PKF is then competitive to the ensemble method where it often needs dozen of members. However in multivariate statistics, the numerical cost of the PKF scales as the square of the number of fields, which is a strong limitation \eg a multivariate assimilation in air quality should consider hundreds of chemical species \cite{Perrot2022}, but in practice, only a few species are assimilated, which makes the PKF interesting for these applications (in CAMS regional air quality production 2.40  \cite{cams}, the univariate 3DVar system of MOCAGE is used for the separated assimilation of ozone, nitrogen dioxide, sulphur dioxide, and fine particulate matter PM2.5 and PM10, following a configuration similar to the one used for MACII detailed by \cite{Marecal2015GMD}).

While it is not the purpose of the present work, we can mention that the analysis step in the PKF consist in a sequential update of the parameters during the assimilation of observations \ie the update of the variance and  of the anisotropy for the VLATcov model after the assimilation of each observation. For instance, at the leading order, the update of the  aspect tensor field due to the assimilation of an in situ observation is written as 
\begin{equation}
    \mathbf{s}^a = \frac{V^a}{V^f}\mathbf{s}^f,
\end{equation}
where $V^a$ is the analysis-error variance field that results from the assimilation of the observation (see update equations Eq.(3) for the variance and Eq.(12) for the aspect tensor in  \cite{Pannekoucke2021T} for details). In the PKF, a sequential processing of batches of observation can be consider to parallelize the assimilation, similarly to the procedure employed in the EnKF analysis step \cite{Houtekamer2001MWR}.

The PKF based on the VLATcov model is illustrated in the next sections for two dynamics which give an explicit form for $\mathcal{M}$ in \Eq{eq:dynamics}.

\subsection{Illustration of the PKF forecast step for simple dynamics}\label{sec:pkf:example}

The transport and the diffusion equations are considered so to detail the dynamics of the variance and the anisotropy for the PKF applied for VLATcov models. Both dynamics play over a 1D periodical domain of coordinate $x$, so that the dynamics is an evolution equation without boundary conditions. The PKF for the transport has been already detailed in previous contributions (see \textit{e.g.} \cite{Pannekoucke2021GMD} ), and is recapped here for self consistency. The PKF for the heterogeneous diffusion equation is original, and extends the derivation of the PKF applied to the homogeneous diffusion equation \cite{Pannekoucke2018NPG}.

\subsubsection{PKF prediction applied on a transport equation}\label{sec:pkf:transport}

The transport equation of a scalar field $c(t,x)$ by a stationary velocity field $u(x)$ is written as
\begin{equation}\label{eq:advection}
    \partial_t c + u \partial_x c =0.
\end{equation}
In this example, and by identification with \Eq{eq:dynamics}, $c$ stands for $\chi$ while $\mathcal{M}(c,\partial c)=-u\partial_x c$.
This kind of equation appears for instance in the prediction of the concentration of a chemical specie as in chemical transport models.

The computation of the PKF dynamics for \Eq{eq:advection} using SymPKF leads to the system
\begin{subequations}
\label{eq:pkf:advection}
\begin{align}
\partial_t c &= - u \partial_x c,\label{eq:pkf:advection:a}\\
\partial_t V_c &= - u \partial_x V_c,\label{eq:pkf:advection:b}\\
\partial_t s_{c,xx} &= - u \partial_x s_{c,xx} + 2 s_{c,xx} \partial_x u,\label{eq:pkf:advection:c}
\end{align}
\end{subequations}
where the anisotropy is represented by the aspect tensor $\mathbf{s}=s_{c,xx}$ in 1D domain. The PKF dynamics \Eq{eq:pkf:advection} is a system of three uncoupled partial derivative equation similar to the one first found by \cite{Cohn1993MWR}.  
This system represents the dynamics of the mean state $\E{c}$, \Eq{eq:pkf:advection:a}, where the expectation operator has been removed for the sake of simplicity ; the transport of the variance, \Eq{eq:pkf:advection:b} ; and the transport of the anisotropy \Eq{eq:pkf:advection:c}, where an additional a source term of anisotropy appears, that is due to the shear by the flow. Compared with an ensemble approach, the PKF approach allows for an understanding of the dynamics and the physics of the uncertainty. 

This example shows that the PKF does not need solving the full state dimension covariance matrix, but only the dynamics of the parameters used in the modeling of the covariance matrix: the variance and the anisotropy in the VLATcov model considered here which evolves from \Eq{eq:pkf:advection:b} and \Eq{eq:pkf:advection:c} respectively. However, the parallelization of the PKF is less obvious than for the parallel computation of an ensemble of forecast, since it requires to parallelize the coupled system \Eq{eq:pkf:advection}. 

Note that the lower script notation $_c$ for $V_c$ and $_{c,xx}$ for $s_{c,xx}$ corresponds to the notation automatically rendered by SymPKF when processing the dynamics \Eq{eq:advection} at a symbolic level. This labelling for the parameters has been introduced when multiple fields are present \eg in multivariate dynamics. While this contribution only address univariate dynamics, the notation is kept here so to facilitate the comparison with the output of SymPKF and also because another important dynamics is discussed: the diffusion equation, which is now presented. 

\subsubsection{PKF prediction applied on a diffusion equation}\label{sec:pkf:diffusion}

The diffusion equation  of a scalar field $f(t,x)$ and of diffusion coefficient $D(x)$,
\begin{equation}\label{eq:diffusion}
    \partial_t f = \partial_x \left(D\partial_x f\right),
\end{equation}
is now considered. This kind of equation appears for instance in the prediction of electron density $f$ of the Earth radiation belts and results from a Hamiltonian formalism  applied on a typical Boltzmann equation, where a Fokker-Planck operator is introduced to evaluate physical interactions responsible for changing particles trapping state \cite{Dahmen2020CPC}. In the radiation belts, the typical spatial coordinates system $x$ in \Eq{eq:diffusion} stands in this case for a combined spatial and physical quantities \eg the energy of the electrons. 
The diffusion equation is also important in the modeling of atmospheric boundary layer where it represents the effect of the turbulence \cite{Stull1988book}.
In this example, and by identification with \Eq{eq:dynamics}, $f$ stands for $\chi$ while $\mathcal{M}(f,\partial f)=\partial_x \left(D\partial_x f\right)$.

The computation of the PKF dynamics for \Eq{eq:diffusion} can be performed using SymPKF. However, because of the second order derivative, the dynamical system
makes appear an unknown term $\E{\varepsilon_f \partial^4_x \varepsilon_f}$,
not determined from $f$, $V_f$ and $s_{f,xx}$ (see \ref{sec:appendixB}). 
An analytical closure has been proposed for 1D domains which states as \cite{Pannekoucke2018NPG}
\begin{subequations}
    \label{eq:P18}
\begin{equation}
    \label{eq:P18:a}
    \E{\varepsilon_f \partial^4_x \varepsilon_f} = 3 g_{f,xx}^{2} - 2 \partial^2_x g_{f,xx}
\end{equation}
when written in metric tensor or
\begin{equation}
    \label{eq:P18:b}
    \E{\varepsilon_f \partial^4_x \varepsilon_f} = \frac{2 \partial^2_x s_{f,xx}}{s_{f,xx}^{2}} + \frac{3}{s_{f,xx}^{2}} - \frac{4 \left(\partial_x s_{f,xx}\right)^{2}}{s_{f,xx}^{3}}
\end{equation}
in aspect tensor, which leads to the PKF dynamics
\end{subequations}
\begin{subequations}
\label{eq:pkf:diffusion}
\begin{align}
\partial_t f &= D \partial^2_x f + \partial_x D \partial_x f,\label{eq:pkf:diffusion:a}\\
\partial_t V_f &= - \frac{2 D V_f}{s_{f,xx}} + D \partial^2_x V_f - \frac{D \left(\partial_x V_f\right)^{2}}{2 V_f} + \partial_x D \partial_x V_f,\label{eq:pkf:diffusion:b}
\end{align}
\begin{multline}\label{eq:pkf:diffusion:c}
\partial_t s_{f,xx} = D \partial^2_x s_{f,xx} + 4 D \\- \frac{2 D \left(\partial_x s_{f,xx}\right)^{2}}{s_{f,xx}} - \frac{2 D s_{f,xx} \partial^2_x V_f}{V_f} +\\
\frac{D \partial_x V_f \partial_x s_{f,xx}}{V_f} + \frac{2 D s_{f,xx} \left(\partial_x V_f\right)^{2}}{V_f^{2}} - 2 s_{f,xx} \partial_x^2 D +\\
2 \partial_x D \partial_x s_{f,xx} - \frac{2 s_{f,xx} \partial_x D \partial_x V_f}{V_f},
\end{multline}
\end{subequations}
where in this dynamical systems, the expected value $\E{f}$ in \Eq{eq:pkf:diffusion:a} is replaced by $f$ for the sake of simplicity.
The dynamics \Eq{eq:pkf:diffusion} makes appear the effect of the transport due the heterogeneity of the diffusion coefficient which implies a flow of velocity $-\partial_x D$, and leads to the same PKF transport dynamics \Eq{eq:pkf:advection} as discussed for \Eq{eq:advection} in the particular case where $u=-\partial_x D$. The other terms in  \Eq{eq:pkf:diffusion} are related to the second-order derivative term $D\partial_x^2 f$, which couples the dynamics of the variance and of the anisotropy.

In term of metric, the closed \Eq{eq:pkf:diffusion} reads as 
\begin{subequations}
\label{eq:pkf:diffusion:g}
\begin{align}
\partial_t f &= D \partial^2_x f + \partial_x D \partial_x f, \label{eq:pkf:diffusion:g:a}\\
\partial_t V_f &= - 2 D V_f g_{f,xx} + D \partial^2_x V_f - \frac{D \left(\partial_x V_f\right)^{2}}{2 V_f} + \partial_x D \partial_x V_f,
\label{eq:pkf:diffusion:g:b}
\end{align}
\begin{multline}\label{eq:pkf:diffusion:g:c}
\partial_t g_{f,xx} = - 4 D g_{f,xx}^{2} + D \partial^2_x g_{f,xx} +\\ \frac{2 D g_{f,xx} \partial^2_x V_f}{V_f} + \frac{D \partial_x V_f \partial_x g_{f,xx}}{V_f} - \frac{2 D g_{f,xx} \left(\partial_x V_f\right)^{2}}{V_f^{2}} +\\ 2 g_{f,xx} \partial_x^2 D + 2 \partial_x D \partial_x g_{f,xx} + \frac{2 g_{f,xx} \partial_x D \partial_x V_f}{V_f}.
\end{multline}
\end{subequations}  

Compared with the use of an ensemble estimation for the error statistics, the dynamics \Eq{eq:pkf:diffusion:g} appears rather complicated, especially the dynamics of the metric \Eq{eq:pkf:diffusion:g:c}. However, this coupled system is simple to solve numerically \eg  considering a finite difference scheme and automatic code generation. Moreover, the numerical cost of \Eq{eq:pkf:diffusion:g} is of the order of three times the cost the original heterogeneous diffusion equation \Eq{eq:diffusion} while it provides an accurate estimation of the error statistics (see numerical experiment in Section~\ref{sec:num:validation:diffusion}), where a large ensemble is needed to reach this accuracy and avoid the sampling noise whose magnitude, for an ensemble of$N_e$ members, scales in $1/\sqrt{N_e}$ from the central limit theorem (CLT). 

Until now, PKF dynamics for the heterogeneous diffusion equation has been evaluated on periodic domain only, while bounded domains are often needed, \eg in radiation belts predictions where the energy of electrons are limited, or in atmospheric boundary layer where the ground is a limit of the domain. The next section addresses how to specify the boundary conditions for the PKF dynamics.

\section{Specification of the boundary conditions for the PKF forecast step}\label{sec:pkfbc}

This section tackles the specification of the boundary conditions for the PKF by considering two usual kind of conditions: the Dirichlet and the Neumann conditions. We consider the particular case of the semi-bounded 1D domain $[0,\infty)$, and focus on the boundary $x=0$. Then we extend to boundary conditions of an arbitrary domain $\Omega$ of frontier $\partial \Omega$.

\subsection{Dirichlet BCs}\label{sec:pkfbc:dirichlet}

A Dirichlet condition at the boundary consists in specifying the value of the fields at $x=0$, that is $\chi(t,x=0)=\chi_0(t)$. 

This conditions is used for the dynamics of the mean in the PKF, but it remains to specify the boundary conditions for the variance and the anisotropy.

Therefore the Dirichlet condition implies that the error field must also verifies a Dirichlet condition \ie $ e(t,x=0)=e_0(t)$. The expectation of the error field at $x=0$ is zero by definition, and of variance $V_0(t) = \E{e_0(t)^2}$. Hence, 
the variance field must also verify a Dirichlet condition \ie $ V(t,x=0)=V_0(t)$.

So for a 1D bounded domain, the Dirichlet condition on the dynamics implies to specify a Dirichlet condition on the variance and on the anisotropy. This result extends for an arbitrary domain $\Omega$ where this time, the boundary conditions for the variance and the anisotropy are Dirichlet conditions on the frontier $\partial \Omega$.

In case where the bounded domain is nested within a larger domain where uncertainty is known from a PKF dynamics, then the variance and the anisotropy at the boundary can be set from the variance and the anisotropy known in the larger domain. When the uncertainty at large scale is featured from an ensemble of forecasts, the statistics at the boundary should be set as the statistics estimated from the ensemble of large scale forecasts at the boundary points \eg for VLATcov models, the variance and the anisotropy can be estimated from the ensemble of large scale forecasts from \Eq{eq:Ve} and \Eq{eq:ge} respectively.

Hence, Dirichlet condition in case of nested models easily extends in 2D and 3D domains where it remains to specify the variance and the anisotropy of the local area model from the variance and the anisotropy of the coupling model.

\subsection{Neumann BCs}\label{sec:pkfbc:neumann}

Neumann conditions at the boundaries are written as null fluxes \ie $\partial_x \chi(t,x=0)=0$. This implies that the error field must also verifies a Neumann condition \ie $\partial_x e(t,x=0)=0$.
Again, we are looking for the boundary conditions for the variance and the anisotropy.

The condition on the variance is deduced from the Taylor expansion of the error at the vicinity of $x=0$ as follows. The expectation of the square of the second order expansion of the error
\begin{equation*}
e(t,\delta x)=e(t,0) + \frac{1}{2}\partial_x^2 e(t,0) \delta x^2+\mathcal{O}(\delta x^3),
\end{equation*}
%
%writes
%$$e(t,\delta x)^2=e(t,0)^2 + \left(e\partial_x^2 e\right)(t,0) \delta x^2+\frac{1}{4}\left(\partial_x^2 %e(t,0) \right)^2 \delta x^4,$$
%which 
leads to the local expansion of the variance $V(t,x)=\E{e^2}(t,x)$,
\begin{equation*}
V(t,\delta x)= V(t,0) + \E{e\partial_x^2e }(t,0) \delta x^2+\mathcal{O}\left(\delta x^4\right).    
\end{equation*}    
As the local Taylor expansion of the variance field at $x=0$, this implies that the first order derivative is null, \ie
\begin{subequations}
\label{eq:neumann}
\begin{equation}\label{eq:neumann-variance}
    \partial_x V (t,0) = 0,
\end{equation}
which means that the condition in variance at the boundary $x=0$ follows a Neumann condition.

For the anisotropy, the Neumann condition on the variance,\Eq{eq:neumann-variance}, implies that the metric tensor $g(t,x)=\E{\left(\partial_x\varepsilon\right)^2}(t,x)$ simplifies as  
$g(t,0) = \frac{1}{V(t,0)}\E{\left(\partial_x e(t,0)\right)^2}$. Then the Neumann condition on $e$, $\partial_x e(t,0)=0$ \ie implies that the condition for the metric is a Dirichlet condition,
\begin{equation}\label{eq:neumann-metric}
    g(t,x=0) = 0.
\end{equation}
\end{subequations}

Note that the Dirichlet condition for the metric \Eq{eq:neumann-metric} is equivalent to an infinite in aspect tensor at the boundary \ie $s(t,0)=+\infty$. Therefore, the PKF formulated in aspect tensor is singular near the boundaries where the error field verifies a Neumann condition. Hence, for a Neumann condition on the error, it is preferable to consider the PKF formulation in variance/metric tensor rather than in variance/aspect tensor (while there is no preference on unbounded domains \eg in periodic domains or flat spaces).

Hence, the Neumann condition on the error for a 1D domain leads, for the PKF dynamics, to a Neumann condition in variance and a Dirichlet condition in metric. This extends to a 2D or 3D domain $\Omega$ where this time the Neumann condition for the variance states as a null flux along the normal direction of the frontier $\partial\Omega$ of the domain. The Dirichlet condition for the metric reads equivalently $g(t,\x)=0$ for $\x\in\partial\Omega$, but a weaker condition could be introduced where the tangential components of the metric at the boundary are not zero (not addressed here).

Now that the boundary conditions for the PKF have been theoretically specified for the Dirichlet and the Neumann conditions, a numerical validation as well as a comparison with the usual EnKF approach is introduced. But to do so, it is necessary to specify an appropriate setting for the boundary of the EnKF, as discussed in the next section.

\section{Methodology for validating the PKF from an EnKF simulations, and Specification of the BCs for EnKF}\label{sec:EnKF:bc}

\subsection{Methodology for the validation of the PKF from an ensemble method}

In the previous sections we specified the PKF equations as well as the boundary conditions it needs depending on the boundary condition for the error at the boundary. 
Note that for linear dynamics as for the transport and the diffusion that are considered here, the dynamics of the error statistics is independent from the dynamics of the mean, so hereafter we will not be considering the dynamics of the mean but rather focusing on the error statistics.
For instance, the PKF dynamics for the error statistics for the transport with a positive wind on a bounded 1D domain and Dirichlet condition in $x=0$, \Eq{eq:pkf:advection} is written as 
\begin{subequations}
\label{eq:pkf-bc:advection}
\begin{align}
\partial_t V_c &= - u \partial_x V_c,\label{eq:pkf-bc:advection:a1}\\
V_c(0,x) &= V_c^0(x),\label{eq:pkf-bc:advection:a2}\\
V_c(t,0) &= b_{V_c}(t),\label{eq:pkf-bc:advection:a3}\\
\partial_t s_{c,xx} &= - u \partial_x s_{c,xx} + 2 s_{c,xx} \partial_x u,\label{eq:pkf:advection:b1}\\
s_{c,xx}(0,x) &= s_{c,xx}^0(x),\label{eq:pkf-bc:advection:b2}\\
s_{c,xx}(t,0) &= b_{s_{c,xx}}(t),\label{eq:pkf-bc:advection:b3}
\end{align}
\end{subequations}    
where $b_{V_c}$ and $b_{s_{c,xx}}$ stands respectively for the error variance and anisotropy time functions prescribed at the inlet of the domain in $x=0$, and where $V_c^0$ and $s_{c,xx}^0$ are the initial condition fields.

Now we are interested to verify that the resulting PKF dynamics, such as \Eq{eq:pkf-bc:advection}, is really able to reproduce the dynamics of the error statistics as an ensemble method would provide. But then, it is necessary to create an ensemble of forecast with an appropriate specification of the boundary error.
Again, for the transport with Dirichlet condition, it remains to perform an ensemble of integration of the dynamics
\begin{subequations}
    \label{eq:enkf:advection}
\begin{align}
\partial_t e_c &= - u \partial_x e_c,\label{eq:enkf:advection:a1}\\
e_c(0,x) &= e_c^0(x),\label{eq:enkf:advection:a2}\\
e_c(t,0) &= b_{e_c}(t),\label{eq:enkf:advection:a3}
\end{align}    
\end{subequations}    
where $b_{e_c}$ and $e_c^0$ stands respectively for the error time function prescribed at the inlet of the domain in $x=0$, and the initial error field.

In real application where the inflow at the inlet is related to physical structures  \eg a mid-latitude vortex entering in a LAM, the error is  continuous and differentiable in time and space meaning this imposes a constraint of coherence between the error field inside the domain and its boundary condition at the inlet. Then, the resulting error statistics should also be continuous and differentiable in time and space: this imposes a constraint of coherence between the variance (or the anisotropy) field inside the domain and its boundary condition at the inlet.

Note that when a covariance model is used in EnKF to perturb boundary conditions, an assumption of stationarity of the error statistics is often encountered, \eg by considering a constant time correlation in an auto-regressive time process \cite{Torn2006MWR} (similar to the one introduced in Appendix~\ref{AppKFBC}). However for physical phenomenons entering in a LAM, the correlation time scale is expected to evolve accordingly to the spatial anisotropy, because the spatial anisotropy is governed by the dynamics and the  observational network \cite{Cohn1993MWR,Bouttier1994MWR,Pannekoucke2021T}. It would therefore be interesting to illustrate how the PKF performs for heterogeneous and non-stationary error statistics at the frontier.

Thus, we consider the following validation framework where the error statistics are given by a prescribed initial condition together with its boundary conditions which leads to smooth evolution of the error statistics, without any constraint of homogeneity and stationarity of the statistics. For instance, when considering the transport, this means that we consider the situation where the error statistics are given by a prescribed initial condition $\left(V_c^0(x),s_{c,xx}^0(x)\right)$ together with its boundary conditions $\left(b_{V_c}(t),b_{s_{c,xx}}(t)\right)$ which leads to smooth fields $\left(V_c(t,x),s_{c,xx}(t,x)\right)$. But, for the ensemble use for the validation, this implies to populate an ensemble of $N_e$ initial conditions $\left({e_c}_k(0,x)\right)_{k\in [1,Ne]}$ together with their boundary conditions $\left({b_{e_c}}_k(t)\right)_{k\in [1,Ne]}$ such that at a given time $t$ and position $x$, the resulting error variance (anisotropy) is $V_c(t,x)$ ($s_{c,xx}(t,x)$). The problem that arises is how to specify the auto-correlation time scale of the boundary perturbation in order to obtain the desired error statistics $\left(V_c(t,x),s_{c,xx}(t,x)\right)$.

Note that ensemble forecasting for the diffusion equation with Neumann boundary conditions corresponds to an initial value problem where each member is integrated from an initial condition that verifies the Neumann conditions, and without additional perturbation of the boundary along the time integration (the metric at the boundary is constant equal to zero).

It results that, the specification of the auto-correlation time scale only concerns the Dirichlet conditions.
In what follows, the specification of the time auto-correlation is first presented for an arbitrary evolution equation, then it is applied for the transport and for the diffusion equation. 

\subsection{Specification of the auto-correlation time-scale of the BC perturbations for ensemble of forecast}

The problem faced here is that, with boundary conditions being time-series, it has an associated time-scale. However, the metric tensor $g_{xx}$ is related to the spatial length-scale of the perturbation as denoted by the index $_{xx}$. In order to specify the boundary condition of the metric tensor field, we need to find an equation linking, on the boundaries, the spatial metric tensor $g_{xx}$ with the time-scale used to generate the perturbation.

Similarly to the spatial metric tensor \Eq{eq:g}, the temporal metric tensor $g_{tt}$ that characterize the auto-correlation of a smooth centered random field, $\eta(t)$, depending on the time, and of variance $V_\eta(t)=\E{\eta(t)^2}$, is defined by
\begin{equation}\label{eq:g_tt_eta}
 \mathbf{g}_{tt}(t)=\E{\partial_t\left(\frac{\eta(t)}{V_\eta(t)}\right)
 \partial_t\left(\frac{\eta(t)}{V_\eta(t)}\right)}.
\end{equation}
This temporal metric tensor is directly related with the time-scale of the perturbation. In 1D $g_{tt} = \frac{1}{L_t^2}$ with $L_t$ the auto-correlation time-scale.

Without loss of generality, the boundary $x=0$ is considered, and the goal is to characterize the temporal metric tensor $g_{tt}(t,x=0)$. If $\eta(t)$ denotes the random error at $x=0$, then by continuity, the error and the random forcing verify $e(t,x=0)=\eta(t)$. Then, it results that
the variances verify $V_\eta(t)=V(t,x=0)$, and the temporal metric tensor reads as
\begin{subequations}
\begin{equation}\label{eq:g_tta}
 \mathbf{g}_{tt,x=0}(t)=\E{\partial_t\varepsilon(t,\x=0)\partial_t\varepsilon(t,\x=0)},
\end{equation}
where $\varepsilon=e/\sqrt{V}$ is the normalized error associated with the spatial error $e$.
While \Eq{eq:g_tta} only holds at the boundary $x=0$, the spatio-temporal smoothness of $e$ implies a link between the temporal metric at the boundary and the spatial metric within the domain, which results from the dynamics of the error \Eq{eq:dyne} at $x=0$: $\partial_t e(t,x=0)=\mathcal{M}(e,\partial e)(t,x=0)$.
In particular, the temporal metric reads as (see \ref{sec:C}) 
\begin{equation}\label{eq:g_ttb}
g_{tt} \underset{x=0}{=} \frac{1}{V}\E{\left(\mathcal{M}(\varepsilon\sqrt{V},\partial (\varepsilon\sqrt{V}))\right)^2} - \frac{1}{4V^2}\left(\partial_t V\right)^2,
\end{equation}
where the terms $\E{\left(\mathcal{M}(\varepsilon\sqrt{V},\partial (\varepsilon\sqrt{V}))\right)^2}$ and $\partial_t V$ can make appear the spatial metric field $g_{xx}$ at $x=0$. 
\end{subequations}

One pitfall is that equation \Eq{eq:g_ttb} may be complicated, and can contain unknown terms such as $\E{\varepsilon_f \partial^4_x \varepsilon_f}$ encountered for the heterogeneous diffusion dynamics in section~\ref{sec:pkf:diffusion}. The next two sub-sections will detail the link between the temporal and the spatial metrics for the transport and for the diffusion. 

\subsection{Dirichlet BC for ensemble forecasting of the positive velocity transport equation}

To illustrate the relation between the temporal and the spatial metric tensor, the
transport equation \Eq{eq:advection} is now considered. 

Following the theoretical derivation of the temporal metric \Eq{eq:g_ttb}, SymPKF is used to derive the relation between the temporal anisotropy tensor of the error and the spatial anisotropy tensor of the error which yields
\begin{equation}\label{eq:adv:metric:dirichlet_1}
g_{c,tt} \underset{x=0}{=} u^{2} g_{c,xx} + \frac{u^{2} \left(\partial_x V_c\right)^{2}}{4 V_c^{2}} + \frac{u \partial_t V_c \partial_x V_c}{2 V_c^{2}} + \frac{\left(\partial_t V_c\right)^{2}}{4 V_c^{2}}.
\end{equation}
This spatio-temporal consistency for the temporal and spatial statistics is difficult to interpret physically without approximations. However, under the assumptions of local homogeneity ($\partial_x V_c=0$) and of stationarity for the variance $\partial_t V_c =0)$, \Eq{eq:adv:metric:dirichlet_1} reads as 
\begin{equation}\label{eq:adv:metric:dirichlet}
        g_{c,tt} \underset{x=0}{=} u^2 g_{c,xx},
\end{equation}
which is physically interpretable since \Eq{eq:adv:metric:dirichlet}, written in time-scale and length-scale, reads as $L_t =\frac{L_x}{u}$: the usual rule relating time and space in a transport. 
Later, the numerical investigation will consider  \Eq{eq:adv:metric:dirichlet} as an approximation of the true time-scale even when assumptions leading to 
\Eq{eq:adv:metric:dirichlet} are not verified. 

Note that \Eq{eq:adv:metric:dirichlet} can be obtained when considering that the dynamics of the variance \Eq{eq:pkf:advection:b} applies at the boundary, leading to replace the trend of the variance by $\partial_t V_c =-u\partial_x V_c $ in \Eq{eq:adv:metric:dirichlet_1} so to obtain \Eq{eq:adv:metric:dirichlet}.

To conclude this paragraph, the ensemble forecasting under Dirichlet boundary conditions and applied to the transport equation, remains to populate an ensemble of boundary perturbations with a prescribed temporal variance and an auto-correlation time scale given by \Eq{eq:adv:metric:dirichlet}.

We proceed in the same way for the diffusion equation.

\subsection{Dirichlet BCs for ensemble forecasting of the diffusion equation}

To continue going towards more and more realistic modeling, the heterogeneous diffusion equation \Eq{eq:diffusion} is now considered to compute the spatio-temporal link \Eq{eq:g_ttb} in the diffusion case.

From a derivation detailed in \ref{sec:appendixA}, the auto-correlation time scale of boundary perturbation can be related to the spatial error correlation length-scale by the proxy
\begin{equation}\label{eq:diff:metric:dirichlet}
        g_{f,tt}(t, x) \approx  3D(x)^2g_{f,xx}(t,x).
\end{equation}
Note that \Eq{eq:diff:metric:dirichlet} is an equality when the variance and the diffusion fields are homogeneous, and when the variance is stationary at the boundary.

Hence, as for the transport, the ensemble forecasting under Dirichlet boundary conditions, and applied to the diffusion equation, remains to populate an ensemble of boundary perturbations with a prescribed temporal variance and an auto-correlation time scale given by \Eq{eq:diff:metric:dirichlet}.

We are now ready to validate the PKF approach from an ensemble validation designed to produce desired error statistics.

\section{Numerical investigation}\label{sec:num:validation}

The goal of the numerical investigation is to validate the PKF on a bounded domain as well as the equations developed in Section~\ref{sec:EnKF:bc}, by comparing the PKF dynamics with an ensemble simulation.

\subsection{Setting for the numerical experiments}

For this investigation three different settings are considered. All experiments take place on a 1D bounded domain $x \in [0,\Lambda]$. For the first one, the transport equation \Eq{eq:advection} is considered with Dirichlet boundary condition at $x=0$ and free boundary at $x=\Lambda$. For the second setting, the heterogeneous diffusion equation \Eq{eq:diffusion} is considered with Dirichlet boundary conditions at both boundaries $x=0$ and $x=\Lambda$. For the third setting, the same diffusion equation is considered but this time with Neumann boundary conditions at $x=0$ and $x=\Lambda$.

The transport and the diffusion being linear, the dynamics of the mean is the same for the PKF and for the EnKF. Hence, without loss of generality, to focus on the validation of the error statistics, the mean state is not considered in the following (the reader can consider the mean state as constant). Then, the ensemble of forecast is equivalent to the forecasts of an ensemble of perturbations $(e_k)_{k\in[1,N_e]}$, with appropriate boundary conditions.

Each time the variance, \Eq{eq:pkf:advection:b} and \Eq{eq:pkf:diffusion:b}, and anisotropy tensor, \Eq{eq:pkf:advection:c} and \Eq{eq:pkf:diffusion:c}, produced by the PKF dynamics are compared with the variance and anisotropy tensor diagnosed from an ensemble of $N_e=6400$ forecasts.
The size of the ensemble considered here, $6400$, is quite unusual compared with operational ensembles that often count dozens of members ; but the reason is that we want to avoid sampling noise to be able to verify that the PKF dynamics of the error statistics reproduces the EnKF considered as the reference. Hence, if a difference appears between the EnKF and the PKF, then it is less likely to be related to sampling noise except when the magnitude of the difference falls under the confident interval deduced from the CLT.

The domain is discretized in $n=241$ grid points and the spatial derivative operator $\partial_x$ is discretized with a centered finite difference scheme leading to a second order of consistency. The temporal discretization scheme varies with each experiment and is detailed in each sections.

\subsection{Application to the transport equation}\label{sec:num:validation:advection}

\begin{figure}
    \centering
    \includegraphics[width=0.65\textwidth]{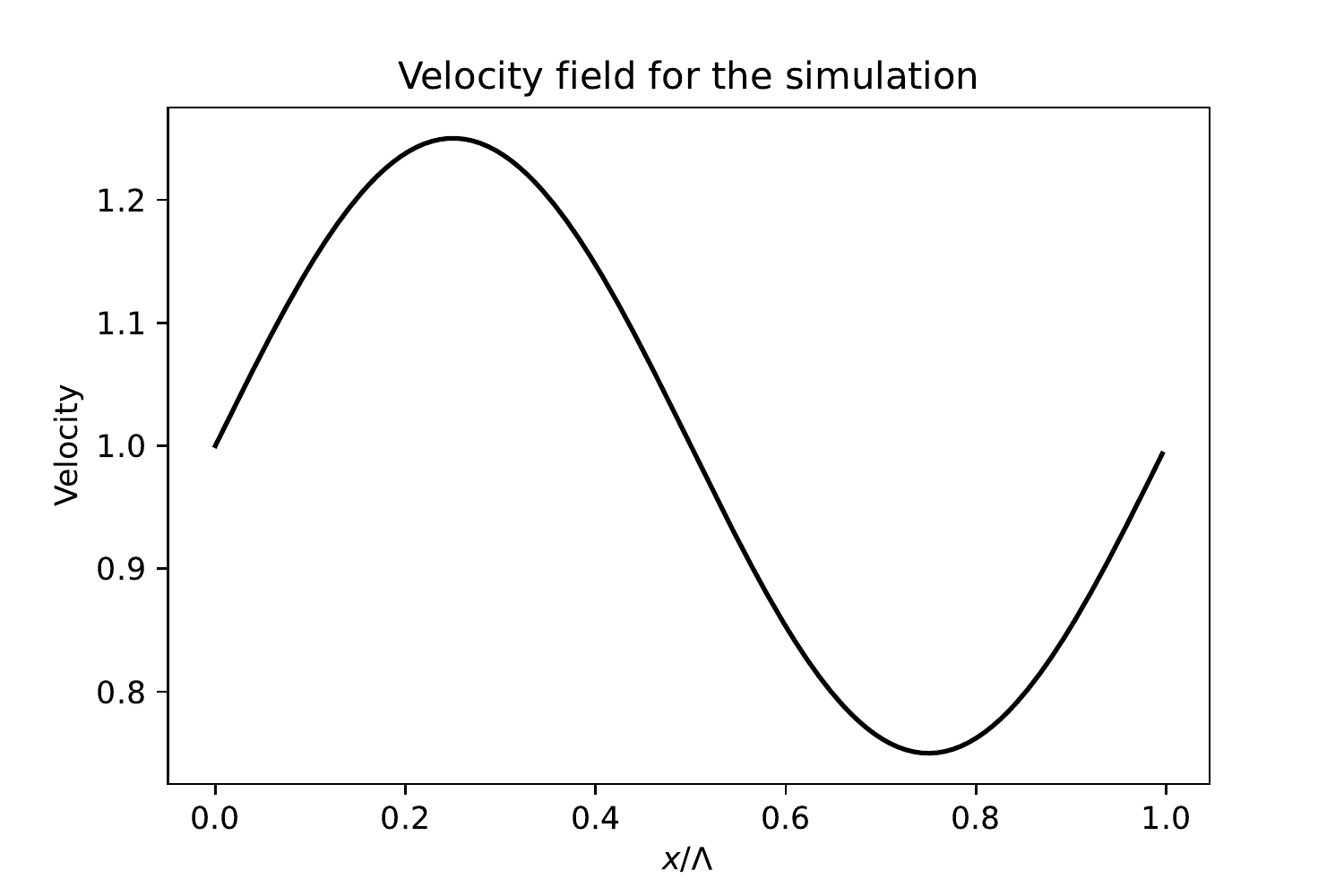}
    \caption{Heterogeneous velocity field considered for the numerical simulation of the transport dynamics.}
    \label{fig:advection:velocity}
\end{figure}

\begin{figure}
    \centering
    \includegraphics[width=\textwidth]{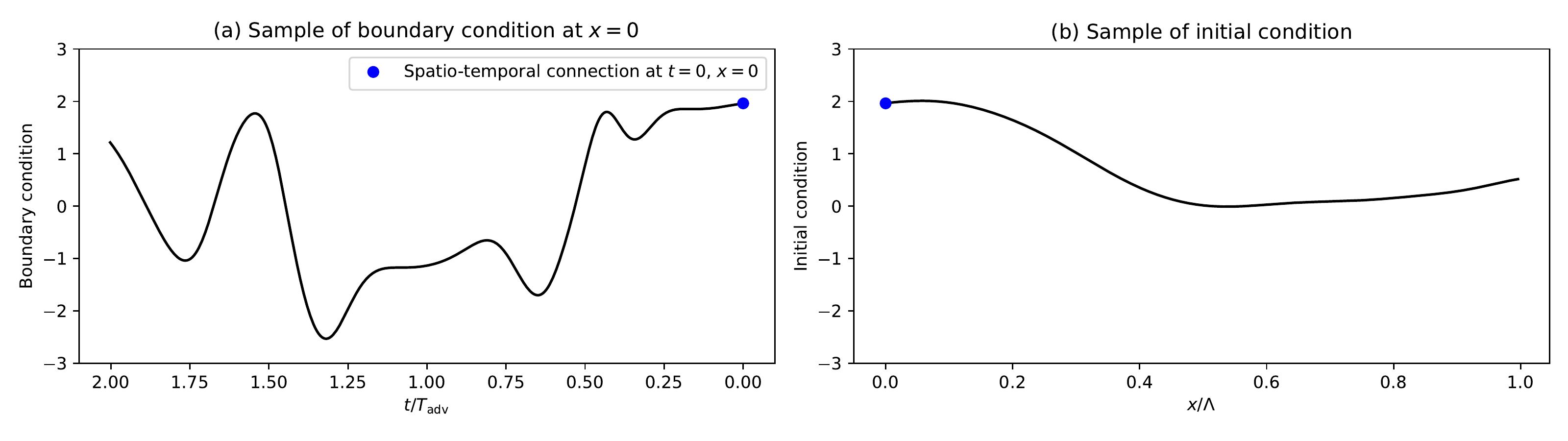}
    \caption{Sample of a generated perturbation split into an initial condition and a boundary condition that are smoothly connected.}
    \label{fig:advection:initial:error:sample}
\end{figure}

\begin{figure}
    \centering
    \includegraphics[width=\textwidth]{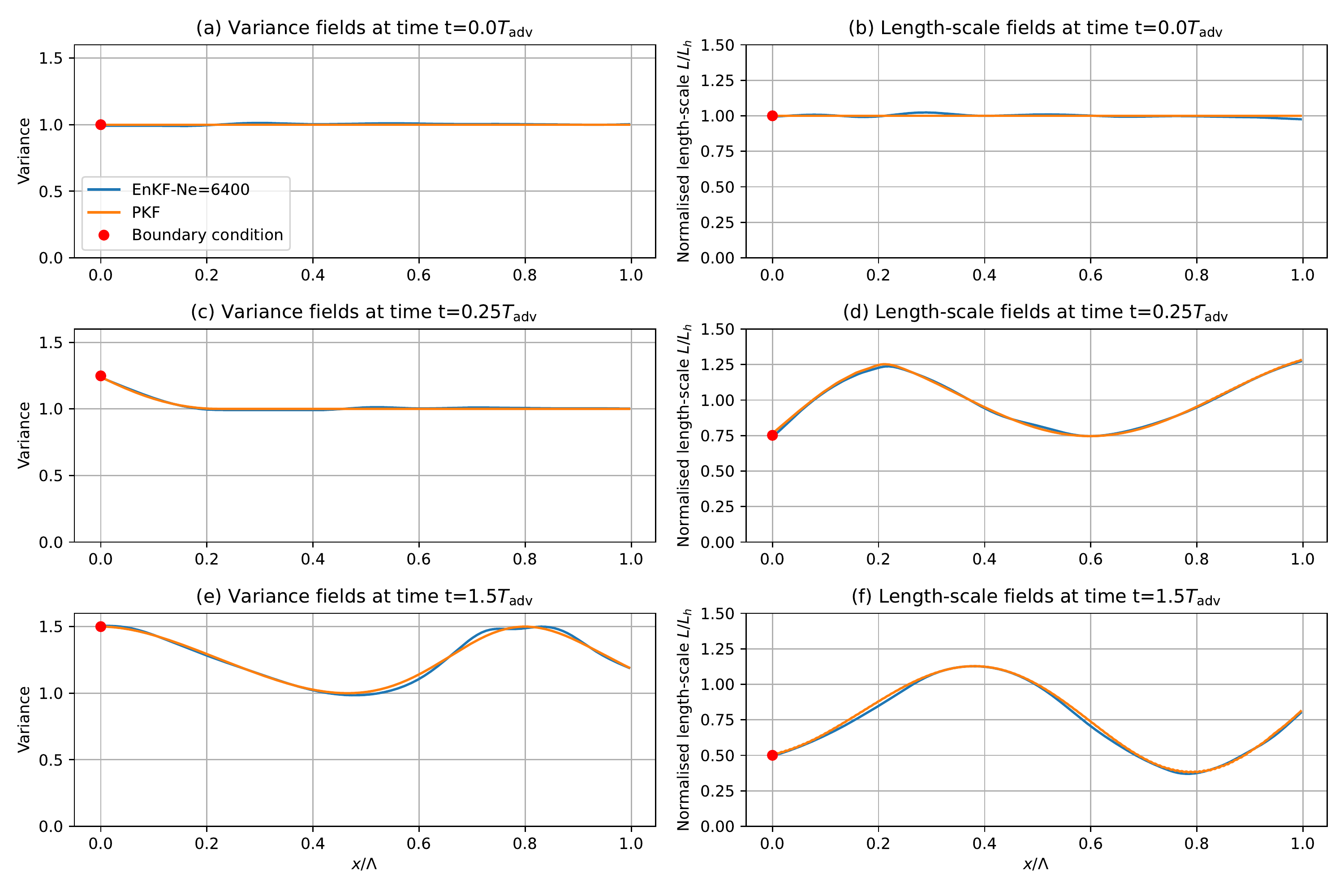}
    \caption{Comparison of the forecast-error variance (left column) and normalized length-scale (right column) fields dynamics for the heterogeneous advection equation on a 1D bounded domain with Dirichlet boundary conditions at $x=0$ and open boundary condition at $x=\Lambda$. The results are shown for times $t=0$, $t=0.25 T_\mathrm{adv}$ and $t=1.5 T_\mathrm{adv}$. 
    }
    \label{fig:advection:dirichlet:snap}
\end{figure}

\begin{figure}
    \centering
    \includegraphics[width=\textwidth]{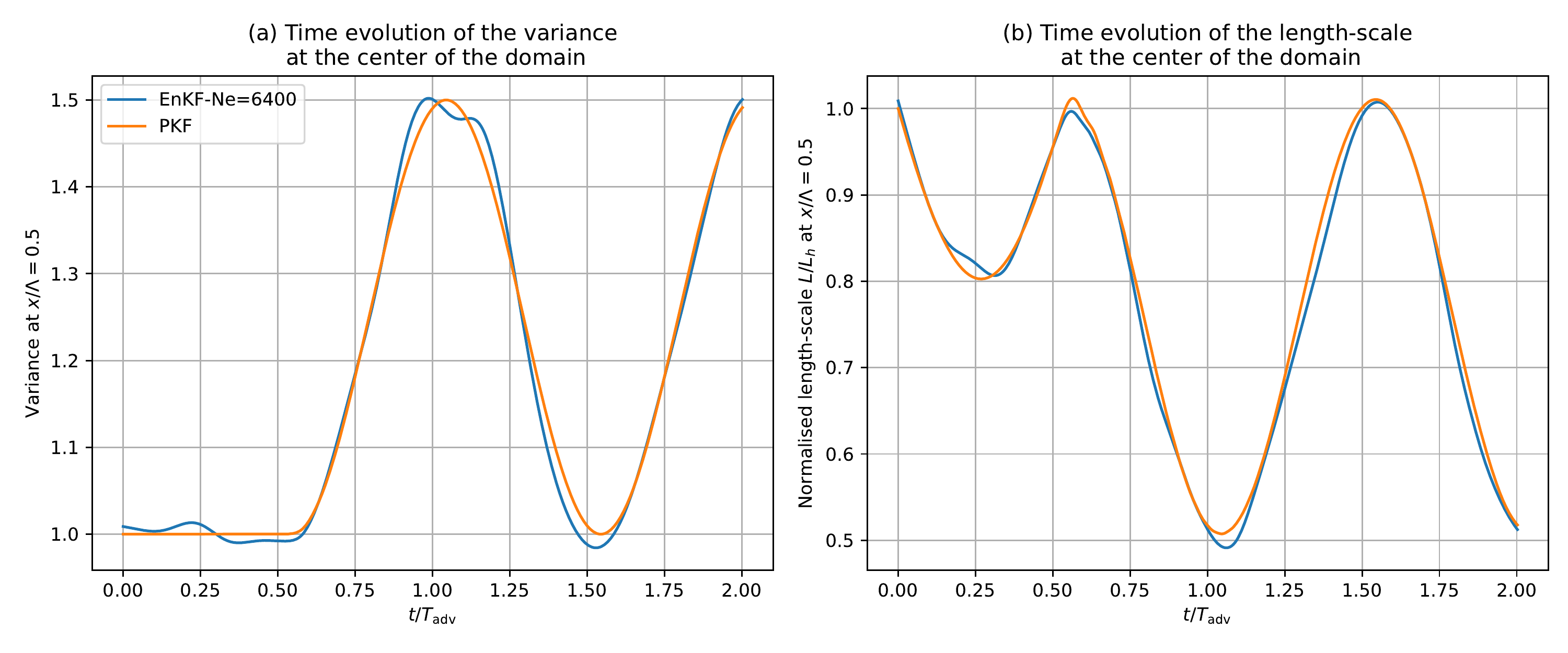}
    \caption{Time evolution of the forecast-error variance (a) and normalized length-scale (b) at $x=0.5\Lambda$, for the advection equation with Dirichlet boundary conditions. }
    \label{fig:advection:dirichlet:time}
\end{figure}

In this experiment setting, the transport equation \Eq{eq:advection} is considered.
The velocity wind for the simulation is set as the heterogeneous stationary field $u(x) = 1 + \frac{1}{4}\sin(\frac{2\pi}{\Lambda}x)$ shown in \Fig{fig:advection:velocity}.

The temporal discretization scheme used for the ensemble simulation as well as the PKF dynamics is a Runge-Kutta scheme of order 4 with a fixed time-step $dt\approx 4. 10^{-3}$. 
The simulation is conducted from time $t=0$ until $t_{end}=2T_{adv}$ with the advection time scale $T_{adv} = \frac{\Lambda}{u_{max}}$ .

In order to generate a coherent set of perturbations for the ensemble simulation \ie an initial condition and a boundary condition that are smoothly connected, an extended domain $[-u(0)t_{end},\Lambda]$ is created from the union of the physical domain $[0,\Lambda]$ and the time window $[0,t_{end}]$ brought back to a virtual physical extension of the domain by multiplying with $-u(0)$.

Then on this extended domain a variance field, $V_0$ and a length-scale field $L_0$ are defined which will be used to generate the perturbations.
For this experiment the fields $V_0$ and $L_0$, that constitute the PKF initial and boundary conditions, are set as follows. The initial variance is set homogeneous and equal to 1 over the physical domain $V_0(t=0,x)=1$ and the boundary variance is set to the periodical function $V_0(t,x=0)=\frac{5}{4} - \frac{1}{4}cos(\frac{2\pi}{T_{adv}}t)$.
Like for the variance, the initial length-scale is set homogeneous and equal to 10\% of the domain length $L_0(t=0,x)=L_h=0.1\Lambda$ and the boundary length-scale is set to the periodical function $L_0(t,x=0)=0.1\Lambda(\frac{3}{4} + \frac{1}{4}cos(\frac{2\pi}{T_{adv}}t))$.

This setting for the variance and the length-scale is chosen so to represent a typical behaviour encountered in numerical weather forecasting, where large scale are more predictable than small scales, which is also the case in radiation belts dynamics forecasting.

Using \Eq{eq:hegauss} and the relation between the length-scale and the anisotropy tensor in 1D, $s_0 = L_0^2$, the covariance matrix, $\mathbf{P}_0=\mathbf{P}(V_0,s_0)$, is defined from which the spatio-temporal perturbations are sampled for each $k$ as $e_k=\mathbf{P}_0^{1/2}\zeta_k$, where $\zeta_k$ is a sample of a centered and normalized Gaussian random vector, and where $\mathbf{P}_0^{1/2}$ stands for the square root matrix of $\mathbf{P}_0$, \ie $\mathbf{P}_0=\mathbf{P}_0^{1/2}\left(\mathbf{P}_0^{1/2}\right)^\mathrm{T}$. The square root $\mathbf{P}_0^{1/2}$ has been computed from the singular value decomposition of the matrix $\mathbf{P}_0$. 

An example of a perturbation sample is presented in \Fig{fig:advection:initial:error:sample} where the temporal evolution $e(t,x=0)$ is shown in panel (a) while the initial condition within the domain, $e(t=0,x)$ is given in panel (b). Note that the time axis in panel (a) has been inverted so to facilitate the understanding. The blue dots corresponds to the value of the sampled error field $e$ at $t=0$ and $x=0$.

The figure \Fig{fig:advection:dirichlet:snap} shows both variance and length-scale fields that are computed from the PKF and the ensemble simulations and compared at three different timestamps. 
The first panels (a) and (b) respectively show the variance and the length-scale normalized by $L_h$ at initial time. As prescribed, the initial variance is homogeneous and equal to $1$. The initial length-scale is also homogeneous and equal to $L_h$.
The panels (c) and (d) present the evolution at time $t=0.25T_{adv}$. As expected the variance (the length-scale) increases (decreases) according to the specified boundary condition close to $x=0$ (red dot). The length-scale is also modified over the whole domain, this is the effect of the velocity gradient of the term $2s_{c,xx}\partial_x u$ in \Eq{eq:pkf:advection:c}.
Since the velocity field is positive, the variance and the length-scale are transported to the right of the domain.
Finally, panels (e) and (f) show the fields at time $t=1.5t_{adv}$. The information injected by the boundary condition at $x=0$ has reached the other side of the domain unscathed for the variance and length-scale fields.

In order to strengthen these results, we show in \Fig{fig:advection:dirichlet:time} the evolution through time of the fields for the middle point in the domain $x=0.5\Lambda$. As expected, the variance in panel (a) remains constant until the information from the boundary condition arrives, where oscillations start, following the prescribed sine shape of the boundary condition shifted in time. In panel (b), the length-scale follows the same kind of dynamic except that the length-scale varies from $t=0$ to $t=0.5T_{adv}$, a variation that is not due to the boundary condition but to the heterogeneity of the wind field. Note that ensemble estimation of the variance and of the length-scale are subject to some sampling noise even with the large ensemble size $N_e=6400$.

Overall, this simulation shows no numerical artifact and the PKF and EnKF forecasts overlap perfectly.
Moreover, the continuous and differentiable error statistics of the EnKF statistics shows that the generated duets of errors for the initial condition and boundary condition have been appropriately specified.

These results validate that the specification of the PKF boundaries proposed in Section~\ref{sec:pkfbc:dirichlet} is correct when applying Dirichlet condition in a transport dynamics. Moreover it also validates the specification of the perturbations \Eq{eq:adv:metric:dirichlet}, introduced in Section~\ref{sec:EnKF:bc}, for the ensemble validation to build prescribed error statistics.

Note that, this example has also shown the ability of the PKF to apply for open boundary condition. 

Now, we validate the PKF boundary conditions applied for a diffusion equation.

\subsection{Application to the diffusion equation}\label{sec:num:validation:diffusion}

In this experiment setting, the heterogeneous diffusion \Eq{eq:diffusion} is considered.
The temporal discretization scheme used for the ensemble simulation is a backward Euler scheme (implicit Euler method) with a fixed time-step $dt_{BE}\approx2. 10^{-4}$. For the PKF dynamics we used a Runge-Kutta scheme of order 4 with a fixed time-step $dt_{RK4}\approx 5. 10^{-6}$. The simulation is performed from time $t=0$ to $t_{end}=1.2T_{\mathrm{diff}}$ with $T_{\mathrm{diff}} = \frac{\Lambda^2}{4D_{max}}$ the time scale of the diffusion of a half-domain.

\begin{figure}
    \centering
    \includegraphics[width=0.65\textwidth]{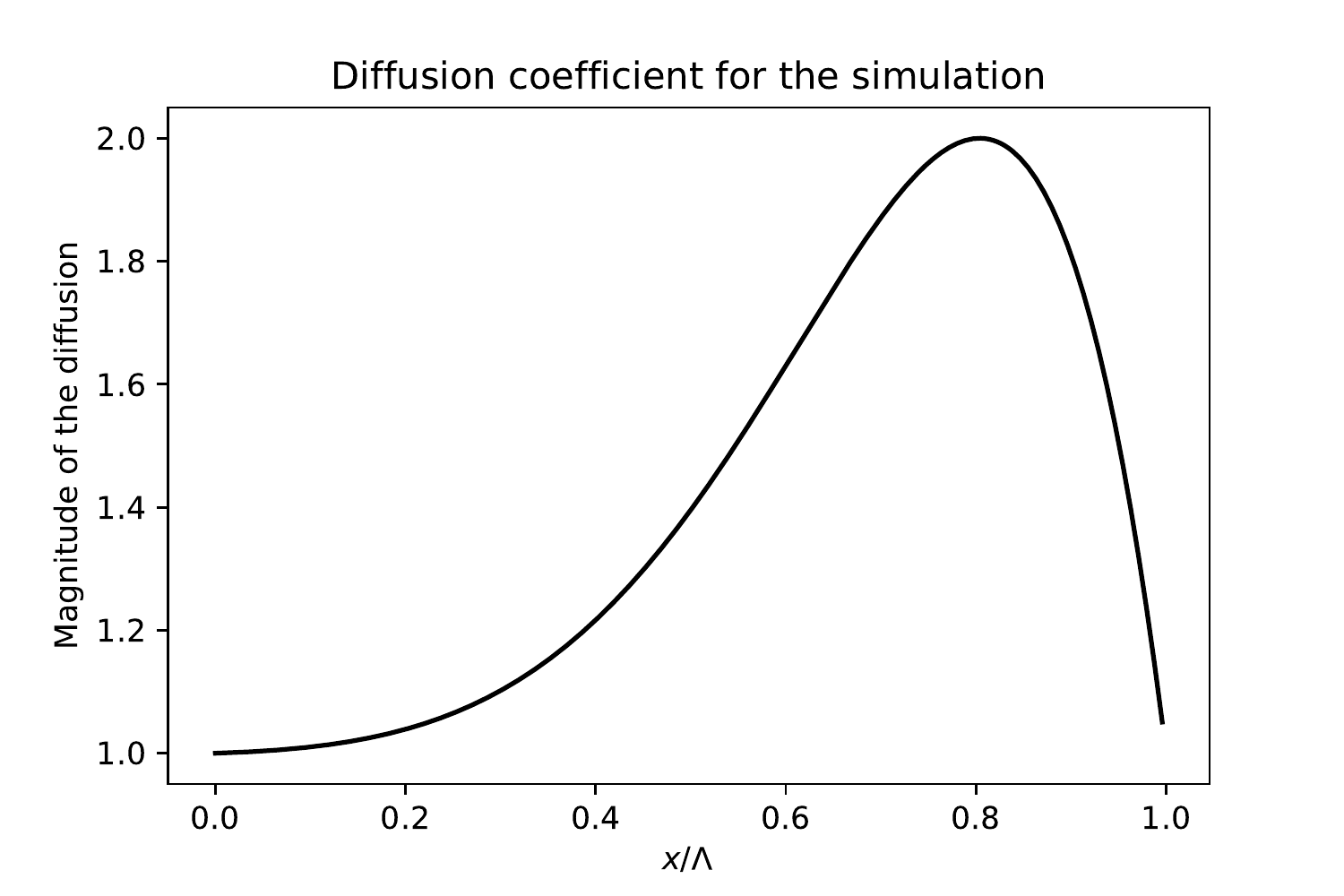}
    \caption{Heterogeneous diffusion coefficient generated for the experiment}
    \label{fig:diffusion:coefficient}
\end{figure}

The diffusion coefficient for the simulation is set as the heterogeneous stationary field $D(x) = 1 + \frac{A}{A_{max}}$ with $A(x) = sin(\pi x)(1+x)^8$ where $A_\mathrm{max} = Max_x A(x)$, and is shown in \Fig{fig:diffusion:coefficient}. This diffusion field reproduces the kind of diffusion encountered in the dynamics of radiation belts in order to evaluate the ability of PKF to solve this problem.

\subsubsection{Dirichlet boundary conditions}\label{sec:num:validation:diffusion:dirichlet}

To generate a coherent set of perturbations for the ensemble simulation, the same technique described in Section~\ref{sec:num:validation:advection} is considered except that both boundaries at $x=0$ and $x=\Lambda$ are subject to Dirichlet conditions. The extended domain considered is $[-\sqrt{D(0)t_{end}},0] \cup [0,\Lambda] \cup [\Lambda,\Lambda+\sqrt{D(\Lambda)t_{end}}]$.

This time, the parameters considered for the simulation and ensemble generation are as follows, the initial variance is set to the linear function $V_0(t=0,x)=1+\frac{3}{\Lambda}x$ and the initial length-scale is set homogeneous and equal to 10\% of the domain length $L_0(t=0,x)=L_h=0.1\Lambda$. For the left boundary condition at $x=0$, the variance and the length-scale are stationary and set equal to $1$ and $L_h$ respectively \ie $V_0(t,x=0)=1$ and $L_0(t,x=\Lambda)=L_h$. For the right boundary condition at $x=\Lambda$, the variance and the length-scale are stationary and set equal to $4$ and $L_h$ respectively \ie $V_0(t,x=\Lambda)=4$ and $L_0(t,x=\Lambda)=L_h$.
From this specification, an ensemble of perturbations has been populated following the same procedure, $e_k=\mathbf{P}_0^{1/2}\zeta_k$, as detailed in Section~\ref{sec:num:validation:advection}. The resulting perturbations are similar to the ones shown in \Fig{fig:advection:initial:error:sample} for the advection, except that there is a right extension of the domain in addition of the left extension for the advection (not shown).

\begin{figure}
    \centering
    \includegraphics[width=\textwidth]{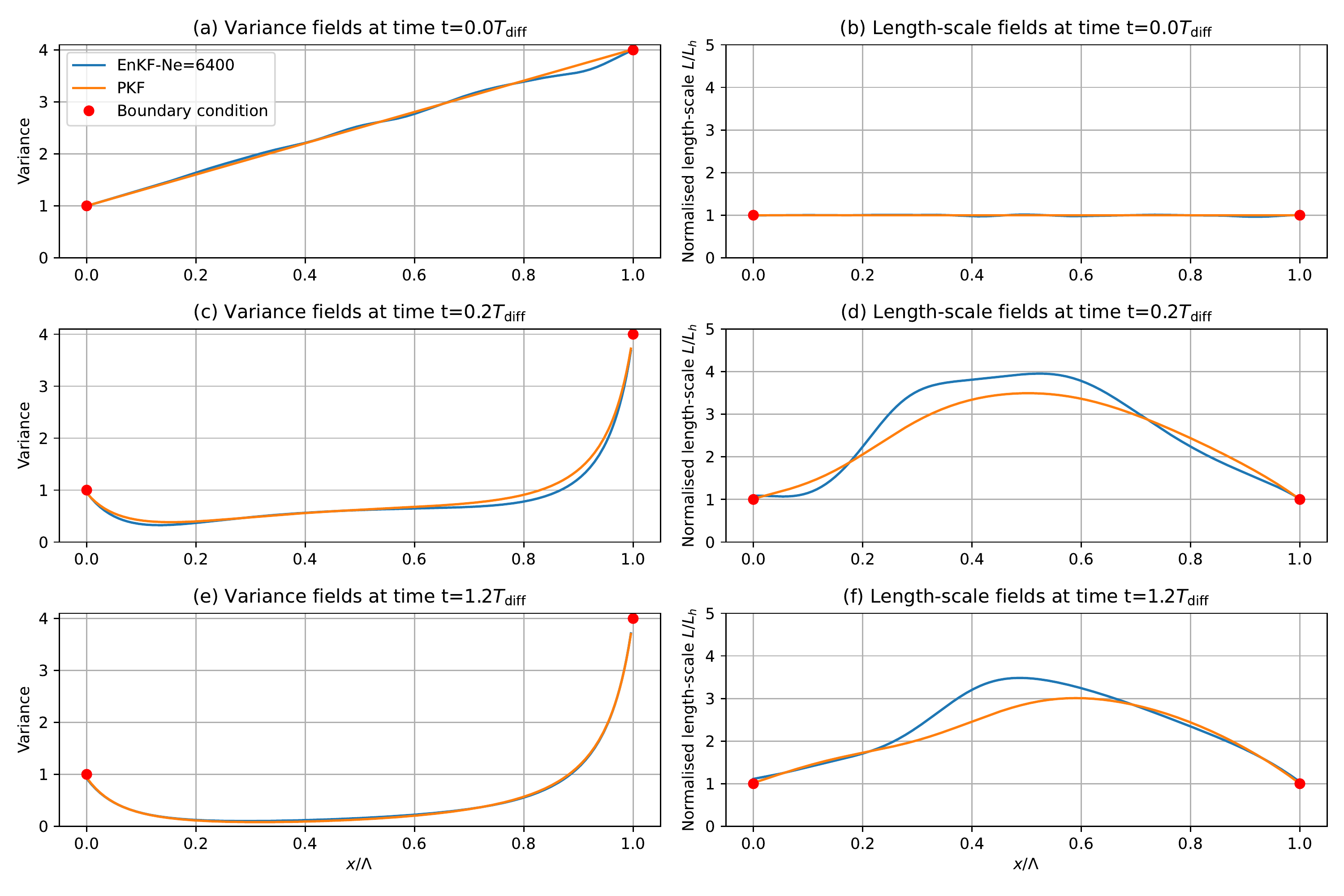}
    \caption{Comparison of the forecast-error variance (left column) and normalized length-scale (right column) fields dynamics for the heterogeneous diffusion equation on a 1D bounded domain with Dirichlet boundary conditions, and shown at times $t=0$, $t=0.2 T_\mathrm{diff}$ and $t=1.5 T_\mathrm{diff}$. 
    }
    \label{fig:diffusion:dirichlet:snap}
\end{figure}

\begin{figure}
    \centering
    \includegraphics[width=\textwidth]{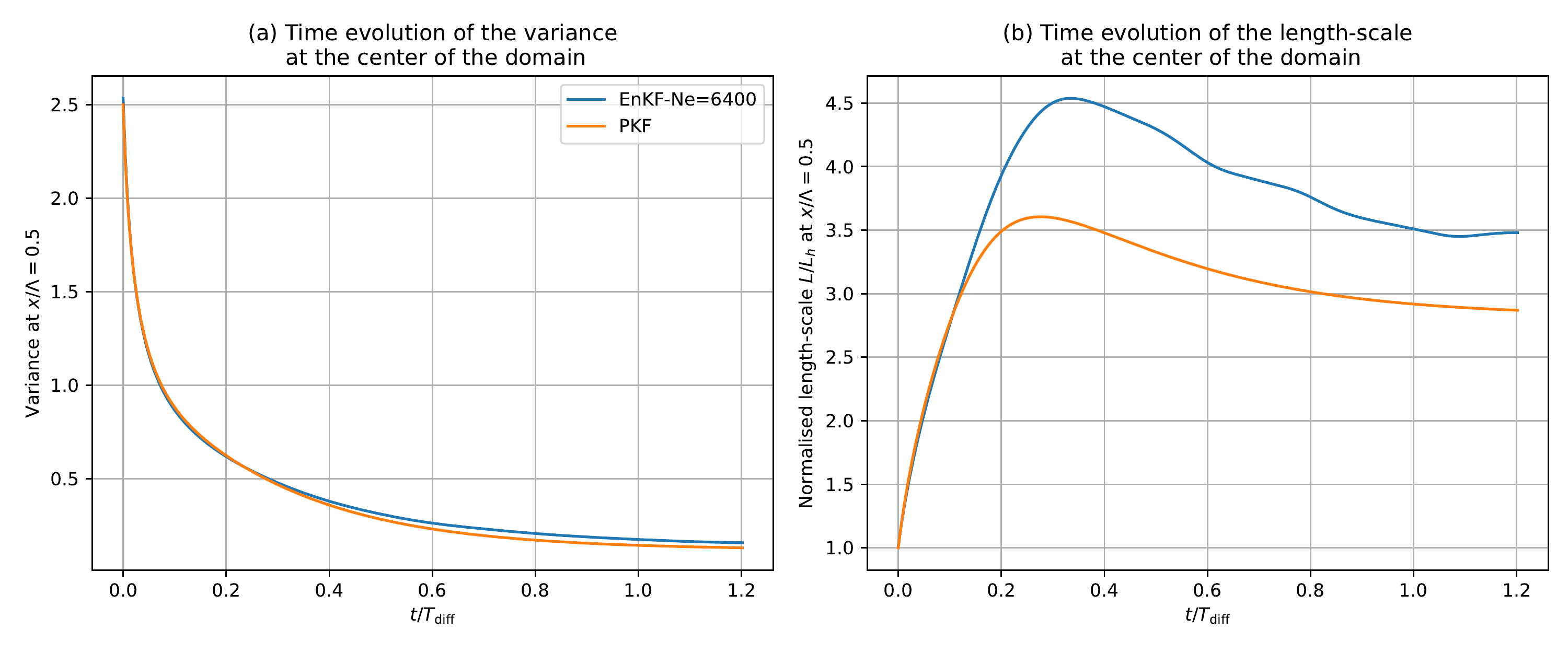}
    \caption{Time evolution of the forecast-error variance (a) and normalized length-scale (b) at $x=0.5\Lambda$, for the diffusion equation with Dirichlet boundary conditions. }
    \label{fig:diffusion:dirichlet:time}
\end{figure}

The comparison between the PKF and EnKF predictions at different time steps are shown in \Fig{fig:diffusion:dirichlet:snap}. The first panels (a) and (b) are coherent with the specification of the initial condition for both the EnKF and the PKF.
Panels (c) and (d) show the evolution of the variance and length-scale at $t=0.2T_{\mathrm{diff}}$. 

Due to physical diffusion, far from the boundaries \eg at the center of the domain, the magnitude of the error is expected to  decrease over time with an attenuation of the variance, while the length-scale should increase; and at the boundaries the uncertainty should remained as specified by the Dirichlet conditions. This is precisely the behaviour observed for both the EnKF and the PKF, at the center of the domain and at the boundaries where the Dirichlet condition imposes fixed values for the variance and the length-scale on both sides of the domain.

However, panel (d) shows a noticeable gap between the length-scale computed by the PKF and the one estimated from the ensemble. This gap can be due to the closure \Eq{eq:P18} but it has a limited impact on the variance field (panel c) which suggests that the PKF prediction of the variance is an accurate proxy for the EnKF estimation.

On the last panels (e) and (f), the variance and the length-scale settle down and the values predicted by the PKF are close to the values computed from the ensemble except for the error observed between the length-scale fields in the middle of the domain. As seen in \Fig{fig:diffusion:dirichlet:time}, the variance and length-scale are close to the permanent regime at $t=1.2T_{\mathrm{diff}}$ showing that the PKF performed well even over a significant time period.

To conclude, this experiment has confirmed the specification of the Dirichlet boundary conditions of Section~\ref{sec:pkfbc:dirichlet} for the PKF applied to a heterogeneous diffusion equation. It has shown the ability of the PKF to accurately approximate the uncertainty dynamics as diagnosed from the EnKF but at a lower cost corresponding the price of two time integrations compared to the $6400$ integrations needed for the ensemble.
Another result is that the simulations also validate the theoretical derivation of the time-scale setting \Eq{eq:diff:metric:dirichlet} needed to obtain a specific length-scale at the boundaries. 

We end the numerical validation by considering the Neumann conditions applied to the heterogeneous diffusion equation.

\subsubsection{Neumann boundary conditions}

\begin{figure}
    \centering
    \includegraphics[width=0.75\textwidth]{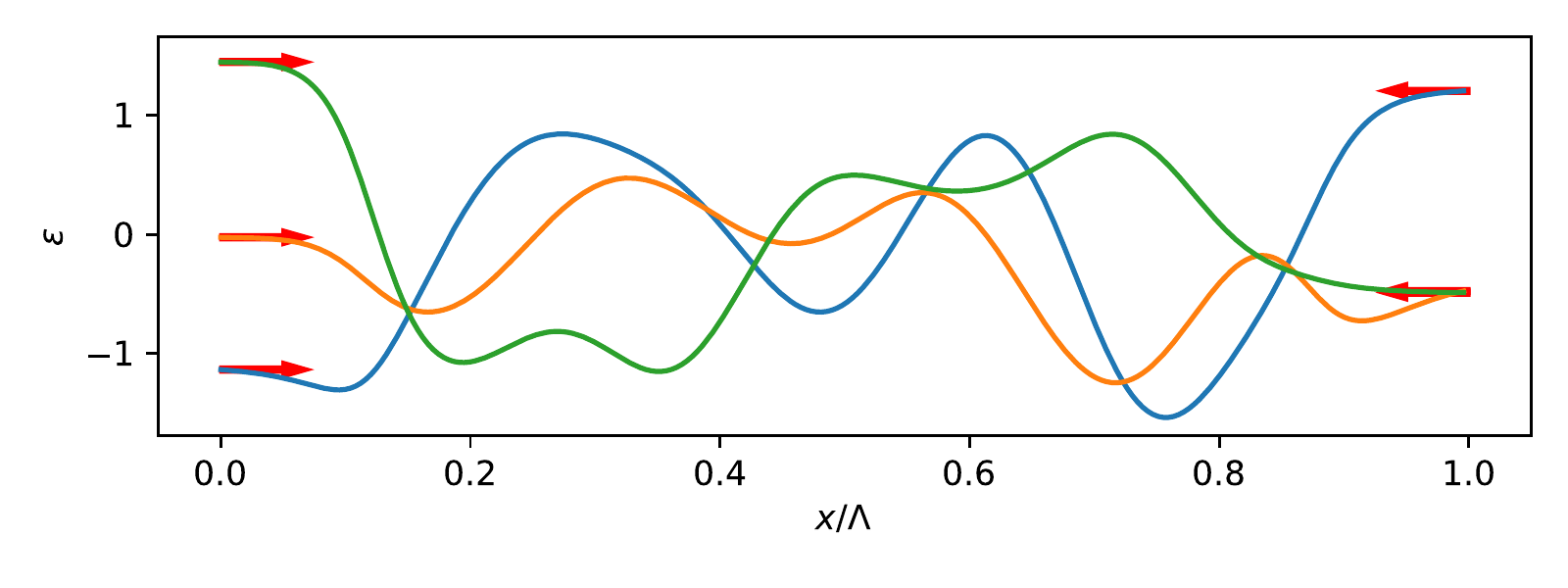}
    \caption{Samples of random error generated as initial condition that verify the Neumann condition at the boundaries $x=0$ and $x=\Lambda$.
    }
    \label{fig:diffusion:neumann:samples}
\end{figure}

\begin{figure}
    \centering
    \includegraphics[width=0.65\textwidth]{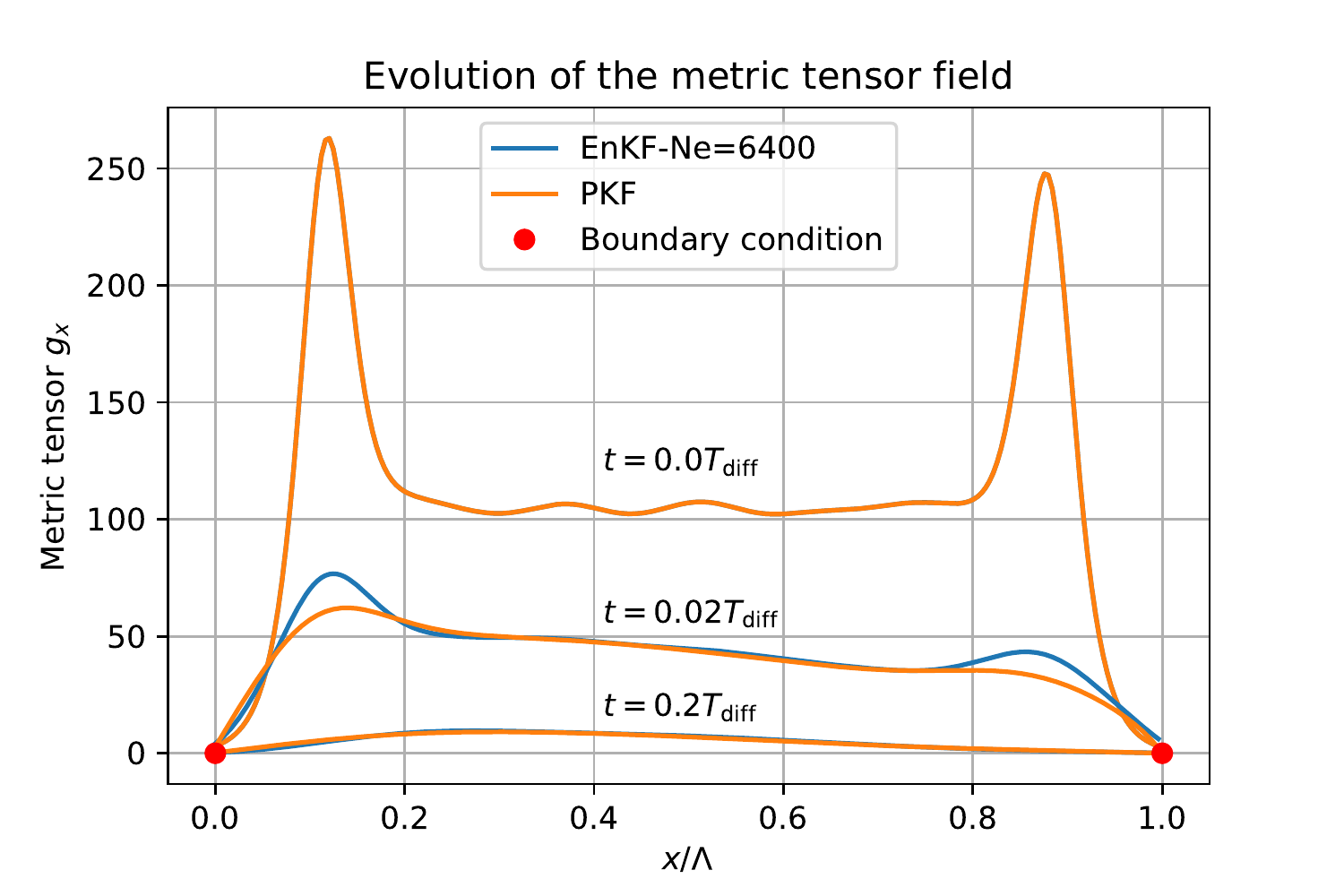}
    \caption{Forecast-error metric field for the heterogeneous diffusion equation on a 1D bounded domain with Neumann boundary conditions, shown at times $t=0$, $t=0.2 T_\mathrm{diff}$ and $t=1.2 T_\mathrm{diff}$. }
    \label{fig:diffusion:neumann:metric}
\end{figure}

\begin{figure}
    \centering
    \includegraphics[width=\textwidth]{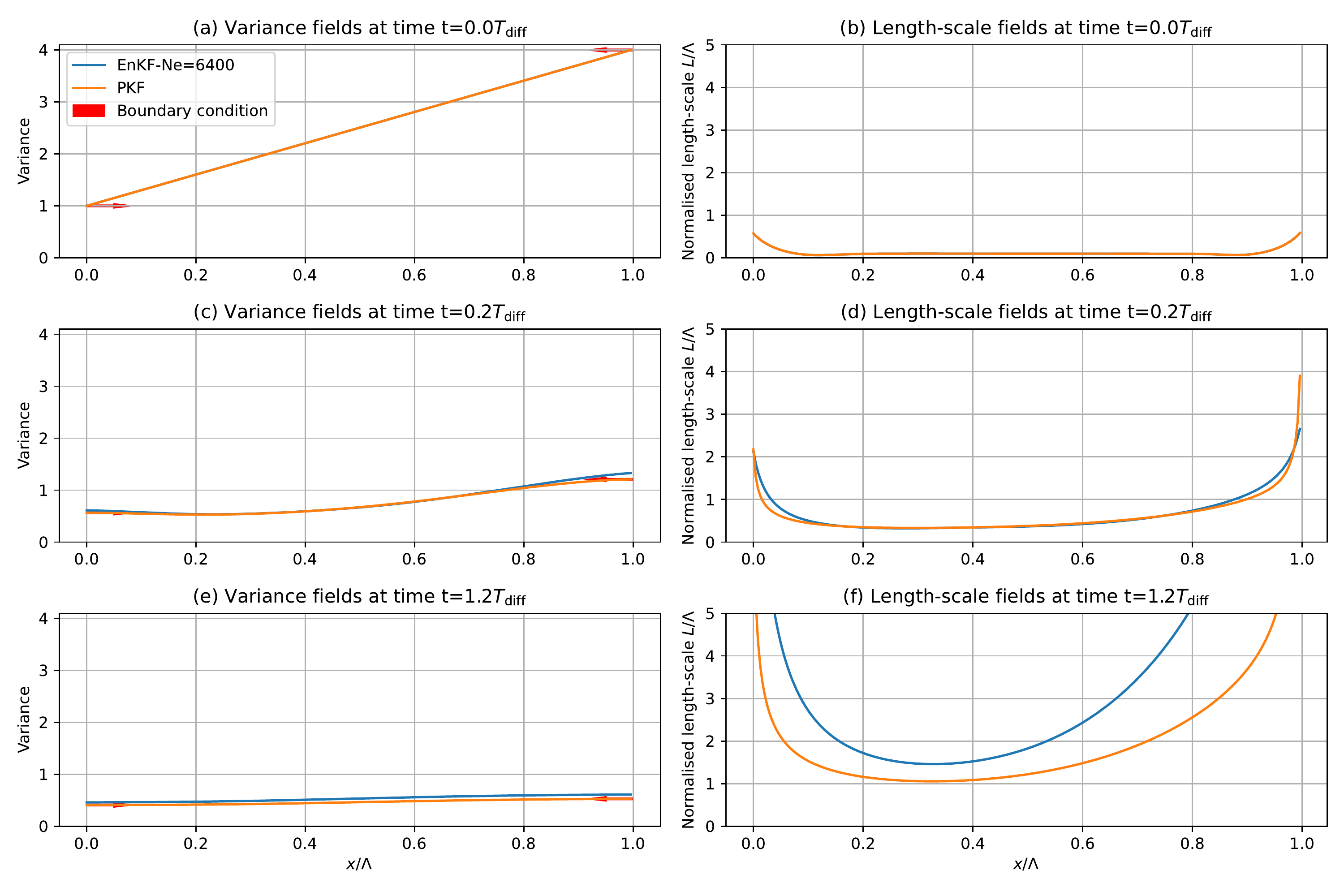}
    \caption{Comparison of the forecast-error variance (left column) and length-scale (right column) fields dynamics for the heterogeneous diffusion equation on a 1D bounded domain with Neumann boundary conditions, and shown at times $t=0$, $t=0.2 T_\mathrm{diff}$ and $t=1.2 T_\mathrm{diff}$. 
    }
    \label{fig:diffusion:neumann:snap}
\end{figure}

\begin{figure}
    \centering
    \includegraphics[width=\textwidth]{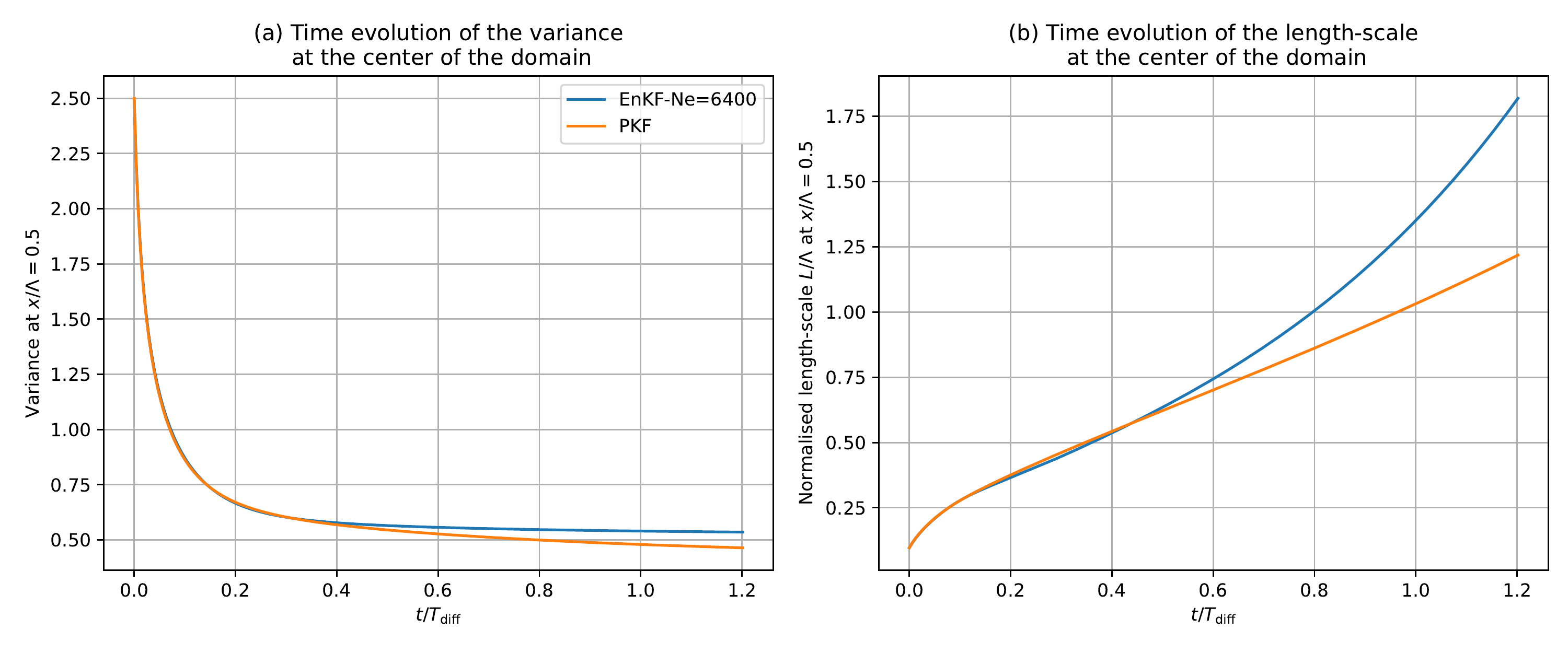}
    \caption{Time evolution of the forecast-error variance (a) and length-scale (b) at $x=0.5\Lambda$, for the diffusion equation with Neumann boundary conditions.}
    \label{fig:diffusion:neumann:time}
\end{figure}

As above mentioned in Section~\ref{sec:EnKF:bc}, compared with the Dirichlet, the Neumann conditions are simulated in an ensemble of forecasts, as an initial condition problem without perturbation at the boundaries. The problem is then to produce an ensemble of initial conditions that verify the Neumann conditions. 

To do so, a covariance model based on a homogeneous pseudo-diffusion equation has been considered \cite{Weaver2001QJRMS}. 
The terminology \textit{pseudo} means that the diffusion is not physical but only a tricky way to create large covariance model as used in variational data assimilation. In particular the square-root covariance $\mathbf{P}_0^{1/2}$ resulting from the integration of the pseudo-diffusion equation reads as the linear operator
    \begin{equation}\label{eq:P0:diffusion}
        \mathbf{P}_0^{1/2} = \mathbf{\Sigma W L},
    \end{equation}
where $\mathbf{L}=e^{\frac{1}{2}\kappa \partial_x^2}$ is the propagator associated with the diffusion equation 
    \begin{equation}\label{eq:P0:pseudo-diffusion}
        \partial_\tau u = \kappa \partial_x^2 u
    \end{equation}
of pseudo-time $\tau$, integrated from $\tau=0$ to $\tau=\frac{1}{2}$, and using Neumann conditions at the boundaries \cite{Mirouze2010QJRMS}; $\mathbf{W}$ is a diagonal normalisation so that $\mathbf{W L}(\mathbf{ W L})^\mathrm{T}$ is a correlation operator ; and $\mathbf{\Sigma}$ is a diagonal matrix of standard deviations, so that the spatial variance field is the linear profile with $V_0(t=0,x=0)=1$ and $V_0(t=0,x=\Lambda)=4$. Note that the pseudo-diffusion coefficient $\kappa$ is related to the length-scale $l$ of the correlation functions as to $\kappa= l^2/2$ \cite{Pannekoucke2008QJRMSa}. For the numerical application, $l_h=0.1\Lambda$.

Again, an ensemble of initial conditions are populated from the square-root \Eq{eq:P0:diffusion}, $e_k=\mathbf{P}_0^{1/2}\zeta_k$, where $\zeta_k$ is a sample of centered Gaussian random vector. \Fig{fig:diffusion:neumann:samples} shows some samples of the normalized error resulting from \Eq{eq:P0:diffusion} \ie $\varepsilon_k=\mathbf{ W L}\zeta_k$. As it is expected, the normalized error are flat at the boundaries (red arrows pointing toward the interior of the domain). The resulting anisotropy diagnosed from the ensemble of initial condition $t=0$ leads to the metric field shown in \Fig{fig:diffusion:neumann:metric} (in blue but super-imposed by the orange line). As expected, far from the boundary, the metric is homogeneous equal to $g=1/l_h^2$ \ie near $x=0.5\Lambda$, but oscillates near the boundaries to reach a value of zero at the boundaries. The oscillations is due to the constraint of symmetry of the covariance matrix \cite{Pannekoucke2018NPG}.

As discussed in Section~\ref{sec:pkfbc:neumann}, for Neumann conditions  the PKF dynamics is solved following its metric formulation, which is given by \Eq{eq:pkf:diffusion:g} for the physical diffusion equation \Eq{eq:diffusion}. For the numerical validation of the Neumann BCs, the initial condition for the PKF is the variance field of linear profile shown in \Fig{fig:diffusion:neumann:snap}-(a) and the experiment metric field diagnosed from the ensemble of initial conditions shown in \Fig{fig:diffusion:neumann:metric} for $t=0$ (orange line, superimposed to the blue line of the EnKF diagnosis).

The PKF dynamics is computed and the results are compared with the ensemble of forecasts of the heterogeneous diffusion equation \Eq{eq:diffusion} and Neumann conditions on both sides. The results are shown in \Fig{fig:diffusion:neumann:metric} for the variance and the length-scale (computed from the inverse of the metric), and in \Fig{fig:diffusion:neumann:metric} for the metric. The results are shown for times of interest selected from the time evolution reproduced in \Fig{fig:diffusion:neumann:time} where a relaxation toward a stationary state of uncertainty appears. 

As expected for a diffusion, the variance decreases along the time, while the length-scale increases. Note that for Neumann condition, the variance at the boundary also decreases while it was constant in the Dirichlet condition. For the ensemble estimation, the length-scales at the boundaries (blue lines in panel (b-d-f) ) are large but finite where it is expected to be infinite: this is due to the numerical estimation of the length-scale deduced from \Eq{eq:ge}, while the metric remains null at the boundaries during the simulation (see \Fig{fig:diffusion:neumann:metric} the red dots). 

Compared with the EnKF diagnosis, the PKF performs well by reproducing the same behaviour of the uncertainty dynamics as for the EnKF, except that the length-scale predicted from the PKF underestimates the length-scale diagnosed from the ensemble. However, the very large length-scale values, larger than the domain size $\Lambda$, as diagnosed from the ensemble is subject to the limitation of the numerical computation of \Eq{eq:ge} for large correlations that can present a positive bias \cite{Pannekoucke2008QJRMS}. Moreover, the large length-scale of the EnKF can also be influenced by model error \cite{Pannekoucke2021NPG}. Because of these limitations, it is not certain that the EnKF reproduces the true dynamics of the uncertainty for these extreme values of the length-scales, while it is considered as the reference. Hence, the discrepancy between the PKF and the EnKF reference, may not be due to a defect of the PKF that could be better than the ensemble estimation here.

To conclude, this experiment has confirmed the specification of the Neumann boundary conditions of Section~\ref{sec:pkfbc:neumann} for the PKF applied to a heterogeneous diffusion equation. It has shown the ability of the PKF to accurately approximate the uncertainty dynamics as diagnosed from the EnKF but at a lower cost corresponding the price of two time integrations compared to the $6400$ integrations needed for the ensemble.

This ends the validation of the specification of the boundary conditions for the PKF. The summary of the results obtained in the paper as well as the perspective of the work are given in the following section of conclusion.

\section{Conclusion}\label{sec:conclusion}

This work contributed to explore the parametric Kalman filter (PKF), that is a recent approximation of the Kalman filter proposed for application in large systems.
The parametric approach investigated here consists to approximate the forecast-error covariance matrix by a covariance model parameterized from the variance and the anisotropy. The anisotropy can be specified in term of metric tensor or its inverse, the aspect tensor, that is the square of the length-scale in 1D domains. The PKF dynamics describes how the mean, the variance and the anisotropy evolve in time, leading to the prediction of the error statistics at a lower numerical cost than the one needed for the full covariance propagation described in the Kalman filter forecast step. 

In this contribution, we proposed how to specify the error statistics at the boundary of a domain when considering a PKF forecast of the uncertainty. We detail here pragmatic solutions for large systems with strong variability at their domain's edge, such as in air quality at regional scale or in radiation belts "weather".

Two kind of boundaries have been considered, the Dirichlet and the Neumann conditions depending on the dynamics. We obtained that a Dirichlet boundary condition for the dynamics implies to specify a Dirichlet conditions for the variance and the anisotropy in the PKF dynamics, no mater how the PKF dynamics is specified in metric or in aspect tensor. Similarly, a Neumann condition implies to specify a Neumann condition for the variance but a Dirichlet condition for the metric. However, for Neuman conditions, the formulation of the PKF in metric is more adapted than the one in aspect tensor, which would require infinite value at the boundary.

%% refactor

Two dynamics of interest for weather forecasting, air quality or radiation belt dynamics have been considered: the transport and the diffusion equation. For these dynamics, the PKF predictions have been validated numerically in a 1D limited area domain, from the comparison to an ensemble estimation \eg by considering heterogeneous and non-stationnary uncertainty at the boundary for Dirichlet conditions. For Dirichlet conditions, the ensemble of forecast has been designed by adding, to each member, a perturbation of the boundary with a varying correlation time-scale, for which we derived analytical formula that depend on the dynamics and introduced to obtain a prescribed spatial anisotropy near the boundary.
For both the advection and the diffusion, the PKF has been shown able to reproduce the uncertainty dynamics diagnosed from the ensemble of forecast.
Note that this result is not obvious, even when considering a linear-Gaussian setting: the PKF based on VLATcov model only approximate the full covariance dynamics from the dynamics of the variance and the anisotropy, that is a crude approximation of the full covariance dynamics.
Hence, it is an encouraging result for the application in air quality (for the transport equation) and in radiation belts monitoring (for the diffusion). For other applications, these results will have to be confirmed by considering non-linear dynamics of interest, and multivariate statistics extension of the PKF, as it is required for numerical weather prediction ; which needs complementary exploration of the PKF beyond the scope of the present contribution and are challenging topics.
Note also that the numerical testbed indirectly validates the auto-correlation time scale formula, and constitutes a contribution to the ensemble methods (while these formula are is not needed for the PKF).

The next step will be to study the BC conditions for domains of larger dimensions, where we expect some changes \eg non-zero components of the metric tensor along the tangential direction to the boundary in Neumann conditions.

We can mention that the dynamics of the uncertainty for bounded domains can be of importance in variational data assimilation or observation targeting applied for local area models, that could be another topic to investigate with the PKF. 

Beyond these challenging topics, we can mention that the results in 1D should already find important applications \eg in the dynamics of uncertainty in the boundary layer for air quality, or wild-land fire predictions ; or in the exploration of the coupling of uncertainty in the coupled atmosphere-ocean system during the alternating integration of the two fluids.

\appendix

\section{Accounting for the uncertainty at the boundary condition in a KF}
\label{AppKFBC}

This section illustrates the KF equation in presence of a boundary, applied to a transport. The KF forecast step is first introduced where the boundary condition appears as a command. Then it is applied when the boundary is perfectly known or uncertain. An illustration of uncertain boundary is proposed when the error is a stationary auto-regressive noise of order 1 ( AR(1) ), for which the formalism can be reworded without command but considering an extended domain with an equivalent model error.

\subsection{KF forecast step for a transport in a limited-area 1D domain}

The transport $\partial_t c+U\partial_x c=0$, of a concentration field $c(t,x)$ at a constant positive velocity $U$, along a 1D limited-area domain $[0,\Lambda]$ of coordinate $x$, with a Dirichlet condition $c(t,x=0)=b(t)$ at the inlet and open condition at the outlet $x=1$, discretized on a regular grid in time $(t_q)_{q\in\mathbb{N}}$ and space $(x_k)_{k\in[0,n-1]}$ (i.e. $x_0=0$) with a time step $\delta t$ and a grid space $\delta x$ defined by $\delta t = \frac{\delta x}{U}$, and when the boundary condition is constant over a time step, the dynamics reads as 
\begin{equation}\label{eq:KFBC}
    c_{q+1}=\mM c_q + \mF b_q,
\end{equation}    
where here, $c_{q+1}$ denote the vector resulting from the discretization of $c(t_q,x)$,   $\mF = [1, 0, \cdots,0]^T$ and  
\begin{equation}
    \mM = \begin{bmatrix}
    0 & 0 & 0 &\cdots&0\\
    1 & 0 & 0 &\cdots&0\\
    0 & 1 & 0&\cdots&0\\
    \vdots&\ddots&\ddots&\ddots&\vdots\\
    &\cdots & 0 & 1&0    
    \end{bmatrix},
\end{equation}
so that the value of the concentration $c(t_{q+1},x_k)$ at a grid point $x_k$ and at a time $t_{q+1}$ is equals to $c(t_{q},x_{k-1})$ providing that $0<k<n-1$ ; at the boundaries the the first component of $c_{q+1}$ is $b_q$ (\textit{i.e.} $c(t_{q+1},0)=b_q$) while the value of $c(t_q,x_{n-1})$. After a long period of time integration, when $t_q>\Lambda/U$,  $c(t_{q+1},x_k) = b_{q-k}$ for any $k\in[0,n-1]$.
Note that, in optimal control, $b_q$ in \Eq{eq:KFBC} plays the role of a command.

From \Eq{eq:KFBC}, it follows that the dynamics of the forecast error is written as 
\begin{equation}\label{eq:KFBC1}
    {e^f_c}_{q+1}=\mM {e^f_c}_q + \mathbf{F}{e_b}_q,
\end{equation}
where $e^f_c$ ($e_b$) denotes the forecast error on $c$ (the error on $b$). As it is common in KF theory, the errors are assumed centered \ie $\E{e^f_c}=\E{e_b}=0$ for any time).

\subsection{Perfectly known boundary condition}

In the case where the boundary condition at the inlet is perfectly known (\ie ${e_b}_q=0$ for any $q$), but the initial condition is uncertain, characterized by Gaussian error distribution of covariance $\mP^a$ ; then the Gaussian uncertainty, characterized by the covariance $\mP^f_q$ at $t_q$, evolves following the Kalman filter forecast step \Eq{eq:KF} that is 
\begin{equation}\label{eq:KFBC2}
\mP^f_{q+1} = \mM\mP^f_q\mM^T.     
\end{equation}
Hence, after one time step ($q=1$), the forecast-error covariance matrix reads as $\mP^f_{1} = \mM\mP^a\mM^T$ ($\mP^f_0=\mP^a$), that is a matrix that looks like 
\begin{equation}\label{eq:KFBC3}
    \mP^f_1 = \begin{bmatrix}
    0 & 0 & 0 &\cdots&0\\
    %0 &  &  &&\\
    0 &  &  &&\\    
    \vdots&& {(\mP^f_0)}^- &&\\   
    0&  &  &&
    \end{bmatrix}
\end{equation}
where ${(\mP^f_0)}^-$ denotes square matrix of dimension $n-1$ whose coefficients are those of the submatrix $\mP^a_{[0:n-2,0:n-2]}$ of $\mP^a$ since $\mP^f_0=\mP^a$. It appears that the variance at the boundary is null, which is coherent when the boundary condition (not necessary constant in time) is perfectly known. After, $q>0$ iterations of \Eq{eq:KFBC2}, $\mP^f_{q+1} = (\mM)^q\mP^f_0(\mM^T)^q$ , the $q^{th}$ first rows and lines of $\mP^f_q$ are null. Since $\mM$ is nilpotent of order $n$ (\ie $\mM^{n-1}\neq0$ and $\mM^{n}=0$), after $n$ iterations, the forecast-error covariance matrix is null, meaning that there is no more error within the domain (independently of the initial uncertainty of the initial condition).

While in this section we considered a perfectly known boundary, this is not the situation encountered in real applications \eg in forecasting at regional scale where uncertainty at large scale should introduce an inflow of uncertainty at the entrance of the domain.

Hence, the inflow of "certainty" that results from the common covariance propagation \Eq{eq:KFBC2} of the Kalman filter forecast step, but used without care, explains the loss of variance observed in ensemble forecasting when no perturbation of the boundary is used to represent the uncertainty \cite{Nutter2004MWR,Nutter2004MWRa}.

The next section provides the correct formulation of the covariance dynamics in presence of uncertainty at the boundary.

\subsection{Uncertain boundary condition}

Now, when considering that the boundary condition is uncertain, the forecast-error covariance matrix evolves as 
\begin{equation}\label{eq:KFBC4}
\mP^f_{q} = \mM\mP^f_q\mM^T + \mF\E{{e_b}_q({e_b}_q)^T}\mF^T+
\mM\E{{e_c^f}_q({e_b}_q)^T}\mF^T + \left(\mM\E{{e_c^f}_q({e_b}_q)^T}\mF^T\right)^T,     
\end{equation}
and reads as 
\begin{equation}\label{eq:KFBC4}
\mP^f_{q} = 
%\mM\mP^f_q\mM^T + 
\begin{bmatrix}
    0 & 0&\cdots&0\\
    0 & &&\\
    %0 &  &  &&\\
    \vdots&& {(\mP^f_q)}^- &\\   
    0&    &&
    \end{bmatrix}+
%\mF\E{{e_b}_q({e_b}_q)^T}\mF^T+
\begin{bmatrix}
    {V_b}_q & 0 &\cdots&0\\
    %0 &  &  &&\\
    0 &  &  &\\    
    \vdots&& {(0)} &\\   
    0&    &&
    \end{bmatrix}+
%\mM\E{{e_c^f}_q({e_b}_q)^T}\mF^T + 
\begin{bmatrix}
    0 & 0 &\cdots&0\\
    %0 &  &  &&\\
    * &  &  &\\    
    \vdots&& {(0)} &\\   
    *&    &&
    \end{bmatrix}+
%\mF\E{{e_b}_q({e_c^f}_q)^T}\mM^T,     
\begin{bmatrix}
    0 & * &\cdots&*\\
    %0 &  &  &&\\
    0 &  &  &\\    
    \vdots&& {(0)} &\\   
    0&    &&
    \end{bmatrix},
\end{equation}
where ${V_b}_q=\E{{e_b}_q({e_b}_q)^T}$ is the variance of the uncertainty at the boundary,
 and where stars, $*$, correspond to \textit{a priori} non zero coefficients which represents the correlations between the uncertainty at the boundary with the forecast-error in the remaining part of the domain $(0,\Lambda]$.
Hence, the first term in \Eq{eq:KFBC4}, $\mM\mP^f_q\mM^T$, is similar to the matrix shown in \Eq{eq:KFBC3}. The other terms in  \Eq{eq:KFBC4} will contribute to fill the first row and column of the matrix: the second term of \Eq{eq:KFBC4}, $\mF\E{{e_b}_q({e_b}_q)^T}\mF^T={V_b}_q\mF\mF^T$,
where $\mF\mF^T$ is the null matrix except for the first coefficient that is one, fills the first coefficient of $\mP^f_{q+1}$. Similarly, the third and the fourth terms of \Eq{eq:KFBC4} will fill the first row and the first column. 

\subsection{Example with a stationary AR(1) boundary error}

As an instructive example, we consider the case where the error is a stationary AR(1) random process, 
${e_b}_{q+1} = a{e_b}_q + \sqrt{1-a^2}\sqrt{V_b}\zeta_q$, where 
$V_b$ denotes the time independent variance of ${e_b}_q$ (independent of $q$) ; $a=\exp{\left(-\delta t/\tau\right)}$ where $\tau$ is the correlation time scale ; and $\zeta = (\zeta_q)_{q\in \mathcal{N}}$ is a process of independent a nd identical independent random variable, where for each $q$, $\zeta_q$ is normal law \ie $\zeta_q\sim\mathcal{N}(0,1)$. The time correlation for this processes is the exponential $cor({e_b}_{q},{e_b}_{q+r})=\exp{\left(-r\delta t/\tau\right)}$, that also reads as $cor({e_b}_{q},{e_b}_{q-r})=\exp{\left(-r\delta t/\tau\right)}$ because of the stationarity of the process (homogeneity in time).

Since after a long period, $c(t_{q+1},x_k) = b_{q-k}$, it follows that ${e_c}(t_{q+1},x_k) = {e_b}_{q-k}$ so that the $k^{th}$ component of the vector $\E{{e_c^f}_q({e_b}_q)^T}$ is 
$cov(e_c^f(t_q,x_k),{e_b}_q)=V_b cor({e_b}_q,{e_b}_{q-k})=V_b cor({e_b}_{q+k},{e_b}_{q}) = V_b cor({e_b}_{q+k},{e_b}_{q}) = V_b \exp{\left(-k\delta t/\tau\right)}$. With $e_c^f(t_{q+1},0)={e_b}_q$, $e_c^f(t_{q+1},x_k)=e_c^f(t_q,x_{k-1})$, and $\delta t = \delta x/U$, then $cov(e_c^f(t_{q+1},x_0),e_c^f(t_{q+1},x_{k}))=V_b\exp{\left(-k\delta x/(U\tau)\right)}$. 
Hence, after a relaxation period, the error covariance matrix is stationnary with a spatial correlation function given as an exponential correlation of length-scale $l=U\tau$.

\subsection{Extension of the space domain for a boundary error modeled by an AR(1)}

When the boundary error is an auto-regressive process or order $1$, it is possible to rewords the dynamics \Eq{eq:KFBC1}, by extending the physical domain with an additional one point corresponding to the boundary condition that will enter to the domain in the future. Focusing on error dynamics, this reads  as follow:

Considering the extended vector $\tilde{e}_q = [{e_b}_q,e^f_q]^T$ of dimension $n+1$, then \Eq{eq:KFBC1} is written as
\begin{equation}\label{eq:KFBCmode}
    \tilde{e}_{q+1}=\widetilde{\mM} \tilde{e}_q + \tilde{\mF}\zeta_q,
\end{equation}
where 
\begin{equation}
    \widetilde{\mM} = \begin{bmatrix}
    a & 0 \\
    \mF&\mM
    \end{bmatrix},
\end{equation}
and $\widetilde{\mF} = [\sqrt{1-a^2}V_b, 0, \cdots,0]^T$.
With this formulation, \Eq{eq:KFBCmode} is similar to the classic formulation of the KF equation in presence of model error (that would be here the second term $\tilde{\mF}\zeta_q$), leading to the error covariance dynamics
\begin{equation}\label{eq:KFBC4}
\widetilde{\mP}^f_{q} = \widetilde{\mM}\widetilde{\mP}^f_q\widetilde{\mM}^T + \widetilde{\mQ},
\end{equation}
where $\widetilde{\mQ} = \widetilde{\mF}\widetilde{\mF}^T$.

This extension of the domain also applies for auto-regressive process of order $m$ where we need to add $m$ points.

\section{Closure of the PKF Dynamics for the diffusion equation} \label{sec:appendixB}    %% Appendix

The computation of the PKF dynamics for the diffusion equation \Eq{eq:diffusion}, with SymPKF, leads to the dynamical system
\begin{subequations}
\label{eq:pkf:diffusion_raw}
\begin{align}
\partial_t f &= D \partial^2_x f + \partial_x D \partial_x f,\\
\partial_t V_f &= - \frac{2 D V_f}{s_{f,xx}} + D \partial^2_x V_f - \frac{D \left(\partial_x V_f\right)^{2}}{2 V_f} + \partial_x D \partial_x V_f,
\end{align}
\begin{multline}
\partial_t s_{f,xx} = 2 D s_{f,xx}^{2} {\mathbb E}\left(\varepsilon_f \partial^4_x \varepsilon_f\right) - 3 D \partial^2_x s_{f,xx} - 2 D +\\
\frac{6 D \left(\partial_x s_{f,xx}\right)^{2}}{s_{f,xx}} - \frac{2 D s_{f,xx} \partial^2_x V_f}{V_f} + \frac{D \partial_x V_f \partial_x s_{f,xx}}{V_f} +\\
\frac{2 D s_{f,xx} \left(\partial_x V_f\right)^{2}}{V_f^{2}} - 2 s_{f,xx} \frac{d^{2}}{d x^{2}} D +\\ 2 \partial_x D \partial_x s_{f,xx} - \frac{2 s_{f,xx} \partial_x D \partial_x V_f}{V_f}
\end{multline}
\end{subequations}
where this time the term $\E{\varepsilon_f \partial^4_x \varepsilon_f}$ is not determined from $f$, $V_f$ and $s_{f,xx}$. This dynamics can be closed considering the closure \Eq{eq:P18}.

\section{Specification of the temporal metric tensor for evolution equations}
\label{sec:C}

This section details the link between the temporal metric \Eq{eq:g_tta}, $\mathbf{g}_{tt}=\E{\partial_t\varepsilon\partial_t\varepsilon}$, and the dynamics of the error.
Since the trend of the normalized error reads as 
\begin{equation}
    \partial_t \varepsilon = \frac{1}{\sqrt{V}}\partial_t e- \frac{1}{2V^{3/2}}e\partial_t V,
\end{equation}
then the temporal metric tensor is written as 
\begin{equation}
g_{tt} = \frac{1}{V}\E{\left(\partial_t e\right)^2} - \frac{1}{V^2}\E{e\partial_t e}\partial_t V+\frac{1}{4V^3}\E{e^2}\left(\partial_t V\right)^2.
\end{equation}
However, we recognize the expression of the variance $V=\E{e^2}$ and its trend, \Eq{eq:dynVa}, so that the temporal metric simplifies as 
\begin{subequations}\label{eqC:gtt}
\begin{equation}\label{eqC:gtta}
g_{tt} = \frac{1}{V}\E{\left(\partial_t e\right)^2} - \frac{1}{4V^2}\left(\partial_t V\right)^2.
\end{equation}
Introducing the trend of the error \Eq{eq:dyne} and by definition of $\varepsilon=e/\sqrt{V}$, the temporal metric reads as 
\begin{equation}\label{eqC:gttb}
g_{tt} = \frac{1}{V}\E{\left(\mathcal{M}(\varepsilon\sqrt{V},\partial (\varepsilon\sqrt{V}))\right)^2} - \frac{1}{4V^2}\left(\partial_t V\right)^2.
\end{equation}
\end{subequations}

\section{Time auto-correlation boundary condition for the diffusion equation}  \label{sec:appendixA}    %% Appendix

The computation of the time auto-correlation metric Leverages on SymPKF. For the diffusion equation, SymPKF leads to 
\begin{multline}
g_{f,tt} = D^{2} {\mathbb E}\left(\varepsilon_f \partial^4_x \varepsilon_f\right) + 2 D^{2} \partial^2_x g_{f,xx} - \frac{D^{2} g_{f,xx} \partial^2_x V_f}{V_f} +\\ \frac{D^{2} \partial_x V_f \partial_x g_{f,xx}}{V_f} + 
\frac{3 D^{2} g_{f,xx} \left(\partial_x V_f\right)^{2}}{2 V_f^{2}} + \frac{D^{2} \left(\partial^2_x V_f\right)^{2}}{4 V_f^{2}}\\ - \frac{D^{2} \left(\partial_x V_f\right)^{2} \partial^2_x V_f}{4 V_f^{3}} +
\frac{D^{2} \left(\partial_x V_f\right)^{4}}{16 V_f^{4}} +
D \partial_x D \partial_x g_{f,xx} +\\ 
\frac{D g_{f,xx} \partial_x D \partial_x V_f}{V_f} + \frac{D g_{f,xx} \partial_t V_f}{V_f} + \frac{D \partial_x D \partial_x V_f \partial^2_x V_f}{2 V_f^{2}}\\
- \frac{D \partial_t V_f \partial^2_x V_f}{2 V_f^{2}} - \frac{D \partial_x D \left(\partial_x V_f\right)^{3}}{4 V_f^{3}} + \frac{D \partial_t V_f \left(\partial_x V_f\right)^{2}}{4 V_f^{3}} +\\ 
g_{f,xx} \left(\partial_x D\right)^{2} + \frac{\left(\partial_x D\right)^{2} \left(\partial_x V_f\right)^{2}}{4 V_f^{2}} - \frac{\partial_x D \partial_t V_f \partial_x V_f}{2 V_f^{2}} +\\
\frac{\left(\partial_t V_f\right)^{2}}{4 V_f^{2}}.
\end{multline}
Considering the analytical closure \Eq{eq:P18} for the unclosed term ${\mathbb E}\left(\varepsilon_f \partial^4_x \varepsilon_f\right)$, the correspondence is written as
\begin{multline}\label{eq:A:g_tt:P18}
g_{f,tt} = 3 D^{2} g_{f,xx}^{2} - \frac{D^{2} g_{f,xx} \partial^2_x V_f}{V_f} + \frac{D^{2} \partial_x V_f \partial_x g_{f,xx}}{V_f} +\\ \frac{3 D^{2} g_{f,xx} \left(\partial_x V_f\right)^{2}}{2 V_f^{2}} + \frac{D^{2} \left(\partial^2_x V_f\right)^{2}}{4 V_f^{2}} - \frac{D^{2} \left(\partial_x V_f\right)^{2} \partial^2_x V_f}{4 V_f^{3}} +\\ \frac{D^{2} \left(\partial_x V_f\right)^{4}}{16 V_f^{4}} + D \partial_x D \partial_x g_{f,xx} + \frac{D g_{f,xx} \partial_x D \partial_x V_f}{V_f} +\\ \frac{D g_{f,xx} \partial_t V_f}{V_f} + \frac{D \partial_x D \partial_x V_f \partial^2_x V_f}{2 V_f^{2}} \\- \frac{D \partial_t V_f \partial^2_x V_f}{2 V_f^{2}} - \frac{D \partial_x D \left(\partial_x V_f\right)^{3}}{4 V_f^{3}} + \frac{D \partial_t V_f \left(\partial_x V_f\right)^{2}}{4 V_f^{3}} +\\ g_{f,xx} \left(\partial_x D\right)^{2} + \frac{\left(\partial_x D\right)^{2} \left(\partial_x V_f\right)^{2}}{4 V_f^{2}} - \frac{\partial_x D \partial_t V_f \partial_x V_f}{2 V_f^{2}} + \frac{\left(\partial_t V_f\right)^{2}}{4 V_f^{2}}.
\end{multline}
The latter expression being quite complex, simplifications are introduced.
First the variance field is assumed locally homogeneous at the boundary \ie $\partial_x V_f(t,x=0)=0$, so that \Eq{eq:A:g_tt:P18} simplifies as 
\begin{multline}\label{eq:A:g_tt:P18:homogeneous}
g_{f,tt} = 3 D^{2} g_{f,xx}^{2} + D \partial_x D \partial_x g_{f,xx} + \frac{D g_{f,xx} \partial_t V_f}{V_f} +\\ g_{f,xx} \left(\partial_x D\right)^{2} + \frac{\left(\partial_t V_f\right)^{2}}{4 V_f^{2}}.
\end{multline}
Then, if the variance is moreover assumed stationary, then \Eq{eq:A:g_tt:P18:homogeneous} becomes
\begin{equation}\label{eq:A:g_tt:P18:homogeneous:stationary}
g_{f,tt} = 3 D^{2} g_{f,xx}^{2} + D \partial_x D \partial_x g_{f,xx}  + g_{f,xx} \left(\partial_x D\right)^{2}.
\end{equation}
Eventually, then the diffusion coefficient field is homogeneous, then 
the spatio-temporal connection between the temporal metric and the spatial metric reads 
\begin{equation}\label{eqA:diff:metric:dirichlet}
        g_{f,tt} =  3 D^2g_{f,xx}.
\end{equation}
While \Eq{eqA:diff:metric:dirichlet} is a particular case, this equality is considered as a proxy for setting the auto-correlation time scale of the boundary perturbation even when the variance and the diffusion fields are heterogeneous.

Note that another expression for the spatio-temporal consistency \Eq{eq:A:g_tt:P18} can be obtained when first considering the dynamics of the variance given by \Eq{eq:pkf:diffusion:b}, leading to replace the trend of the variance by $\partial_t V_f = - 2 D V_f g_{f,xx} + D \partial^2_x V_f - \frac{D \left(\partial_x V_f\right)^{2}}{2 V_f} + \partial_x D \partial_x V_f$, so that \Eq{eq:A:g_tt:P18} simplifies as 
\begin{multline}
g_{f,tt} = 2 D^{2} g_{f,xx}^{2} + \frac{D^{2} \partial_x V_f \partial_x g_{f,xx}}{V_f} +\\ \frac{D^{2} g_{f,xx} \left(\partial_x V_f\right)^{2}}{V_f^{2}} + D \partial_x D \partial_x g_{f,xx} +\\ \frac{2 D g_{f,xx} \partial_x D \partial_x V_f}{V_f} + g_{f,xx} \left(\partial_x D\right)^{2},
\end{multline}
from which the assumption of local homogeneity at the boundary \ie $\partial_x V_f(t,x=0)=0$, 
leads to
\begin{equation}
g_{f,tt} = 2 D^{2} g_{f,xx}^{2} + D \partial_x D \partial_x g_{f,xx} + g_{f,xx} \left(\partial_x D\right)^{2}.    
\end{equation}
When the diffusion field is constant, then the time auto-correlation metric 
is related to the space auto-correlation metric by 
\begin{equation}\label{eqA:diff:metric:dirichlet_bis}
        g_{f,tt} =  2D^2g_{f,xx},
\end{equation}
which is a different result from \Eq{eqA:diff:metric:dirichlet}.

It is not clear whether the appropriate consistency should be given by \Eq{eqA:diff:metric:dirichlet} or \Eq{eqA:diff:metric:dirichlet_bis} \ie if it is 
right to replace the trend of the variance \Eq{eq:pkf:diffusion:b} in the consistency relation \Eq{eq:A:g_tt:P18}.

From numerical experiment, it appears that setting the time auto-correlation of boundary perturbation with \Eq{eqA:diff:metric:dirichlet} in the EnKF is in agreement with the PKF results. This suggests that taking into account the trend of the variance would lead to a kind of over-specification of the boundary condition for the diffusion equation in an EnKF approach.

\section{Open Research}
V1.0 of the \emph{Boundary conditions for the parametric kalman filter forecast} software used to compute and analyze the numerical experiments presented in this paper is preserved at 10.5281/zenodo.7193985 and developed openly at https://github.com/opannekoucke/pkf-boundary. \cite{git-pkf-boundary}

%%%%%%%%%%%%%%%%%%%%%%%%%%%%%%%%%%%%%%%%%%%%%%%

%%%%%%%%%%%%%%%%%%%%%%%%%%%%%%%%%%%%%%%%%%%%%%%

\section*{Acknowledgements}
The authors would like to thank François Rogier (ONERA) for fruitful discussions.
M. Sabathier PhD Thesis is supported by CNES grant 51/19168.
O. Pannekoucke is supported by the French national program LEFE/INSU grant "Multivariate Parametric Kalman Filter" (MPKF).

\includegraphics[scale=0.1]{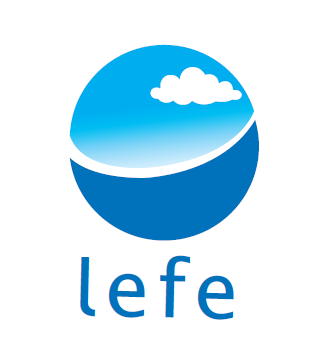}

\bibliographystyle{plain} 
\bibliography{library.bib}

\end{document}